\documentclass[prd,twocolumn,superscriptaddress,nofootinbib,floatfix]{revtex4-2}

\usepackage{amsmath,amsthm,amssymb,booktabs,dcolumn}
\usepackage{graphicx}
\usepackage{txfonts}
\usepackage{ulem}

\usepackage[colorlinks=true,
            citecolor=blue,
            anchorcolor=red,
            menucolor=red,
            linkcolor=red,
            filecolor=red,
            runcolor=red,
            urlcolor=blue]{hyperref}

\usepackage{orcidlink}

\graphicspath{{Figures/}}

\newcommand{\lzu}{\affiliation{School of Physical Science and Technology, Lanzhou University, Lanzhou 730000, China}}
\newcommand{\lctp}{\affiliation{Lanzhou Center for Theoretical Physics,
Key Laboratory of Theoretical Physics of Gansu Province,\\
Key Laboratory of Quantum Theory and Applications of MoE,\\
Gansu Provincial Research Center for Basic Disciplines of Quantum Physics, Lanzhou University, Lanzhou 730000, China}}
\newcommand{\Isotopes}{\affiliation{MoE Frontiers Science Center for Rare Isotopes, Lanzhou University, Lanzhou 730000, China}}
\newcommand{\CSR}{\affiliation{Research Center for Hadron and CSR Physics, Lanzhou University and Institute of Modern Physics of CAS, Lanzhou 730000, China}}
\newcommand{\ahstu}{\affiliation{School of Electrical and Electronic Engineering, Anhui Science and Technology University, Bengbu 233000, China}}

\begin{document}

\title{Revisiting charmonium hybrid spectroscopy}

\author{Ri-Qing Qian\orcidlink{0000-0002-5352-1243}}\email{qianrq@lzu.edu.cn}
\lzu\lctp\CSR
\author{Bing Chen\orcidlink{0000-0001-7561-833X}}\email{chenbing@ahstu.edu.cn}
\ahstu
\author{Xiang Liu\orcidlink{0000-0001-7481-4662}}\email{xiangliu@lzu.edu.cn}
\lzu\lctp\Isotopes\CSR

\begin{abstract}
Hadrons with explicit gluonic degrees of freedom, such as charmonium hybrids, are key to understanding nonperturbative behavior of strong interaction, yet they remain experimentally elusive. Within a constituent gluon model treating the hybrid as a $c\bar{c}g$ three-body system with a transverse electric gluon, we predict the masses of the lightest hybrid multiplet ($J^{PC}=1^{--}, 0^{-+}, 1^{-+}, 2^{-+}$) to lie in the range $4.19$--$4.32$ GeV. By analyzing their decay patterns, we provide specific, experimentally testable signatures to guide the search for these exotic states at current and future facilities.
\end{abstract}

\date{\today}
\maketitle

\section{Introduction}

Understanding the role of gluonic degrees of freedom in hadron structure is a central challenge in nonperturbative QCD. Hybrid mesons, which contain an explicit gluon excitation in addition to a quark-antiquark pair, offer a unique window into this dynamics, distinct from conventional mesons and baryons.

In the light-flavor sector, the search for hybrids with exotic quantum numbers $J^{PC}=1^{-+}$ has yielded several candidates, including the $\pi_1(1400/1600)$~\cite{Aoyagi:1993kn,E852:1997gvf,E852:1998mbq,CrystalBarrel:1998cfz,IHEP-Brussels-LosAlamos-AnnecyLAPP:1988iqi,Khokhlov:2000tk,VES:2001rwn,Baker:2003jh,COMPASS:2009xrl,CLEO:2011upl,JPAC:2018zyd,CrystalBarrel:2019zqh}, $\pi_1(2015)$~\cite{E852:2004gpn,E852:2004rfa}, and the more recent $\eta_1(1855)$~\cite{BESIII:2022iwi,BESIII:2022riz}. Lattice QCD and phenomenological models predict masses for light hybrids that are broadly consistent with these observations. However, recent detailed analyses of their decay patterns suggest that the $\pi_1(1600)$ and $\eta_1(1855)$ cannot both be explained as hybrid states~\cite{Zhang:2025xee,Ma:2025cew}, challenging a simple interpretation and leaving the identification of light-flavor hybrids as an open question.

Heavy-flavor hybrids, while comparatively less explored~\cite{Barnes:1995hc,Zhu:1998sv,Kalashnikova:2008qr,Chen:2013zia,Kalashnikova:2016bta,Oncala:2017hop,Farina:2020slb}, offer significant theoretical advantages. The large charm quark mass provides a higher energy scale, making perturbative treatments of the gluon more reliable. This facilitates both spectral calculations---where the gluon can be modeled as a constituent particle in a $c\bar{c}g$ system---and decay analyses, where the quark-gluon coupling is more tractable. Furthermore, in addition to the usual selection rules for hybrid decays~\cite{Page:1996rj}, heavy-quark spin symmetry imposes further constraints on decay patterns. These features make the experimental identification of charmonium hybrids particularly promising.

In this work, we perform a systematic study of charmonium hybrid mesons within a constituent gluon model. The hybrid is treated as a $c\bar{c}g$ three-body system with a transverse electric (TE) mode gluon, from which we calculate the spectrum. We then evaluate their decay dynamics using the QCD coupling between quarks and gluons, identifying the dominant open-charm channels. Based on these results, we provide specific predictions to guide experimental searches for these elusive states at facilities such as BESIII, Belle II, and the forthcoming Super Tau-Charm Facility.

\section{The charmonium hybrid mass spectrum}

\subsection{Gluonic degrees of freedom in hybrids}

Unlike conventional mesons, hybrid mesons contain explicit gluonic excitations. Modeling this gluonic degree of freedom is a key theoretical challenge. Several complementary approaches exist, including the flux tube model~\cite{Isgur:1984bm,Barnes:1995hc,Close:1994hc}, the QCD string model~\cite{Kalashnikova:2002tg}, and the constituent gluon model~\cite{Horn:1977rq,Swanson:1998kx}. In the flux tube picture, the gluonic field in the strong-coupling limit condenses into a collective, string-like configuration between the quarks, approximated by a chain of massive "beads." The constituent gluon model, by contrast, treats the gluon as a dynamical quasi-particle analogous to a constituent quark.

These pictures are connected in a specific limit. When the flux tube is approximated by a single bead, the hybrid simplifies to a three-body system of quark, antiquark, and bead. This description aligns with both the QCD string model~\cite{Kalashnikova:2002tg} and the constituent gluon model~\cite{Swanson:1998kx}. This single-bead approximation has been shown to accurately reproduce the conventional quarkonium spectrum~\cite{Barnes:1995hc}. Building on this, a three-body hybrid model was recently developed in Ref.~\cite{Chen:2025pvk}, where the massive bead is interpreted as a constituent gluon.

Central to this description are the quantum numbers of the gluon. While a free gluon carries $j_g^{PC}=1^{--}$, the confined environment alters its effective quantum numbers. The bag model predicts that the lowest gluonic excitation inside a confining cavity is a transverse electric (TE) mode with $j_g^{PC}=1^{+-}$~\cite{Barnes:1982tx,Chanowitz:1982qj}. This is consistent with lattice QCD, which finds that the lightest hybrid multiplet corresponds to a hybrid containing a $j_g=1$ TE gluon~\cite{HadronSpectrum:2012gic,Cheung:2016bym}. 
%In this work, we therefore adopt a constituent gluon as a TE mode.

The angular momentum state of such a gluon can be decomposed into transverse and longitudinal components (see Appendix.~\ref{app:TEM_gluon}), with the physical gluon being transverse. Following Ref.~\cite{Farina:2020slb}, the TE and transverse magnetic (TM) gluon states are written compactly as:
\begin{equation}\label{eq:gluon_angular}
    \boldsymbol{Y}_{j_g,m_g}^{(\xi)}(\theta,\phi) = \sum_{\mu,\lambda} \sqrt{\frac{2j_g+1}{4\pi}} D_{m_g,\mu}^{j_g*}(\phi,\theta,0)\chi_{\mu,\lambda}^{(\xi)} \boldsymbol{\epsilon}(\lambda,\hat{\boldsymbol{n}}) \,,
\end{equation}
where $\hat{\boldsymbol{n}}$ is the unit vector in the $(\theta,\phi)$ direction. The index $\xi=-1$ denotes a TE gluon and $\xi=+1$ a TM gluon. The coefficients $\chi_{\lambda,\mu}^{(\xi)}$ are given by
\begin{equation}
    \chi_{\lambda,\mu}^{(-)} = -\frac{\lambda}{\sqrt{2}}\delta_{\lambda,\mu},\qquad \chi_{\lambda,\mu}^{(+)} = \frac{1}{\sqrt{2}}\delta_{\lambda,\mu}.
\end{equation}
Using the Wigner rotation property $D^{j_g}_{m_g,\mu}(\phi+\pi,\pi-\theta,0)=(-1)^{j_g}D^{j_g}_{m_g,-\mu}(\phi,\theta,0)$, the parity of the TE/TM gluon is
\begin{equation}
    P_g = \xi (-1)^{j_g}\,.
\end{equation}

When this gluon couples with a quark-antiquark pair to form a hybrid meson, the total parity and charge conjugation are
\begin{equation}
    P = \xi(-1)^{j_g+l_{c\bar{c}}+1}\,,\qquad C = (-1)^{l_{c\bar{c}}+S_{c\bar{c}}+1}\,,
\end{equation}
where $l_{c\bar{c}}$ and $S_{c\bar{c}}$ are the orbital angular momentum and total spin of the $c\bar{c}$ pair, respectively. 
For the lowest hybrid multiplet, the $c\bar{c}$ pair is in an $l=0$ state. In Table~\ref{tab:hybrid_states}, we present the corresponding quantum numbers of the lowest hybrid multiplet with a $j_g=1$ TE- or TM-mode gluon. For the TE gluon, the $1^{--}$ hybrid state has $c\bar{c}$ spin $S=0$, while the $0^{-+}$, $1^{-+}$, and $2^{-+}$ states also arise from the $S=1$ configuration. Among these, the $1^{-+}$ state carries exotic quantum numbers not accessible to conventional $c\bar{c}$ mesons. For the TM gluon, the four states have quantum numbers $1^{+-}$, $0^{++}$, $1^{++}$, and $2^{++}$, none of which are exotic.

In general, we expect that hybrid with a TM gluon will have a higher mass than the corresponding TE gluon hybrid. In the na\"ive bag model, the TM-mode gluon have an energy approximately 0.5 GeV above the TE-mode~\cite{Barnes:1982tx}.
According to lattice QCD results~\cite{Juge:2002br}, the TM gluon may correspond to the higher excited gluon field, such as the $\Sigma_g^{+\prime}$ and $\Pi_g$ fields. The hybrid spectrum obtained by the lattice QCD~\cite{HadronSpectrum:2012gic,Cheung:2016bym} indicates that the lowest hybrid multiplet have quantum numbers consistent with hybrid states containing a $j_g=1$ TE gluon, while the next excited hybrid multiplet consistent with an $l_{c\bar{c}}=1$ state coupled to a $j_g=1$ TE gluon.
Therefore, hybrid mesons containing the TM gluon may be difficult to produce in experiments. In this work, we therefore restrict our attention to $c\bar{c}g$ states containing the TE gluon.

\subsection{Mass spectrum}

\begin{table}
    \caption{The quantum number of lowest charmonium hybrid multiplet with TE-mode and TM-mode constituent gluon with $j_g=1$.}\label{tab:hybrid_states}
    \begin{ruledtabular}
        \begin{tabular}{ccccc}
            $J^{PC}$ (TE gluon) & $1^{--}$ & $0^{-+}$ & $1^{-+}$ & $2^{-+}$ \\
            \hline
            $J^{PC}$ (TM gluon) & $1^{+-}$ & $0^{++}$ & $1^{++}$ & $2^{++}$ \\
            \hline
            $(l_{c\bar{c}},S_{c\bar{c}})$ & $(0,0)$ & $(0,1)$ & $(0,1)$ & $(0,1)$ 
        \end{tabular}
    \end{ruledtabular}
\end{table}

We now calculate the mass spectrum of charmonium hybrids using the constituent gluon model, treating the hybrid as a $c\bar{c}g$ three-body system. The constituent gluon carries quantum numbers $J^{PC}=1^{+-}$, corresponding to a TE mode with $j_g=1$. Following our previous work~\cite{Chen:2025pvk}, the non-relativistic Hamiltonian for the system is
\begin{equation}
    H =   \sum_{i=c,\bar{c},g }\left(m_i+\frac{p_i^2}{2m_i}\right) + \sum_{i=c,\bar{c}}V_{FT}(r_{ig}) + \sum_{i\neq j}V_{\text{OGE}}(r_{ij}) + c_H \,,
\end{equation}
where $V_{FT}$ is the confining flux-tube potential between the quark/antiquark and the gluon, and $V_{\text{OGE}}$ is the one-gluon exchange potential between the $c$ and $\bar{c}$. Following the flux tube model~\cite{Isgur:1984bm,Barnes:1995hc}, the confining potential is linear:
\begin{equation}
    V_{FT}(r_{ig}) = -\frac{3}{4}C_{ig} \,b\,r_{ig} \, ,
\end{equation} 
with color factor $C_{cg}=C_{\bar{c}g}=-3/2$. The one-gluon exchange potential is
\begin{equation}
    V_{\text{OGE}}(r_{ij}) = C_{ij}\left( \frac{\alpha_s}{r_{ij}} - \frac{8\pi}{3}\frac{\alpha_s}{m_im_j}\tilde{\delta}(r_{ij})\boldsymbol{s}_i\cdot\boldsymbol{s}_j \right) \,,
\end{equation}
where $C_{c\bar{c}}=1/6$ and $\tilde{\delta}(r)=(\sigma/\sqrt{\pi})^3e^{-\sigma^2 r^2}$ smears the contact interaction. The spin-dependent part is treated perturbatively.

In Jacobi coordinates, the spin-independent Schr\"odinger equation becomes
%\begin{widetext}
\begin{eqnarray}
    &&\left[ \frac{\boldsymbol{p}_\rho^2}{2m_\rho} + \frac{\boldsymbol{p}_\lambda^2}{2m_\lambda} + \sum_{i=c,\bar{c}}\left(-\frac{3}{2}\frac{\alpha_s}{r_{ig}}+\frac{9}{8}br_{ig}\right)+\frac{1}{6}\frac{\alpha_s}{\rho}+c_H\right]\psi_H(\rho,\lambda)\nonumber\\ &&= E_H\psi_H(\rho,\lambda) \,,
\end{eqnarray}
%\end{widetext}
where $\boldsymbol{\rho}$ is the relative coordinate between $c$ and $\bar{c}$, $\boldsymbol{\lambda}$ is the coordinate between the gluon and the $c\bar{c}$ center of mass, and the reduced masses are $m_\rho = m_c/2$ and $m_\lambda = 2m_c m_g/(2m_c+m_g)$.

The wave function $\psi_H \equiv \psi_{n_\lambda,n_\rho}^{l_\lambda m_\lambda;l_\rho,m_\rho}(\boldsymbol{\rho},\boldsymbol{\lambda})$ is an eigenstate of orbital angular momenta $l_\rho$ and $l_\lambda$. To preserve the constituent gluon's $1^{+-}$ quantum number assignment, we restrict to the ground state in the $\lambda$-mode ($l_\lambda=0$) and solve variationally using a simple harmonic oscillator (SHO) trial wave function:
\begin{equation}\label{eq:SHO}
    \psi_{0,0}^{00;00}(\boldsymbol{\rho},\boldsymbol{\lambda}) = \frac{\beta_\rho^{3/2}}{\pi^{3/4}}e^{-\frac{\beta_\rho^2\rho^2}{2}} \frac{\beta_\lambda^{3/2}}{\pi^{3/4}}e^{-\frac{\beta_\lambda^2\lambda^2}{2}}.
\end{equation}

The spin-dependent potentials yield mass shifts for the four $l=0$ states:
\begin{align}
    \Delta m_{1^{--}} &= \frac{3}{4}\zeta,\quad
    \Delta m_{0^{-+}} = -2\epsilon -\frac{1}{4}\zeta,\\
    \Delta m_{1^{-+}} &= -\epsilon-\frac{\zeta}{4},\quad
    \Delta m_{2^{-+}} = \epsilon-\frac{\zeta}{4},
\end{align}
where
\begin{eqnarray}
    \epsilon &=& \langle \psi_H|\frac{4\alpha_s}{\sqrt{\pi}}\frac{\sigma^3}{m_cm_g}e^{-\sigma^2 r_{ig}^2}|\psi_H\rangle,\\
    \zeta &=& \langle \psi_H|\frac{4\alpha_s}{9\sqrt{\pi}}\frac{\sigma^3}{m_c^2} e^{-\sigma^2\rho^2}|\psi_H\rangle.
\end{eqnarray}
The mass splitting between the $1^{--}$ and the $0^{-+},1^{-+},2^{-+}$ states arises from $c\bar{c}$ spin-dependent interactions, while the splittings among the $0^{-+}$, $1^{-+}$, and $2^{-+}$ states originate from spin-dependent quark-gluon interactions.

Using the variational wave function, these matrix elements evaluate to
\begin{eqnarray}
    \epsilon &=& \frac{4\alpha_s}{\sqrt{\pi}}\frac{\sigma^3}{m_cm_g}\frac{8\beta_\lambda^3\,\beta_\rho^3}{(4\sigma^2\beta_\rho^2+4\beta_\lambda^2\,\beta_\rho^2+\beta_\lambda^2\sigma^2)^{3/2}},\\
    \zeta &=& \frac{4\alpha_s}{9\sqrt{\pi}}\frac{\sigma^3}{m_c^2}\frac{\beta_\rho^3}{(\beta_\rho^2+\sigma^2)^{3/2}}.
\end{eqnarray}

\begin{figure}[htbp]
  %\centering
  \includegraphics[width=8.6cm]{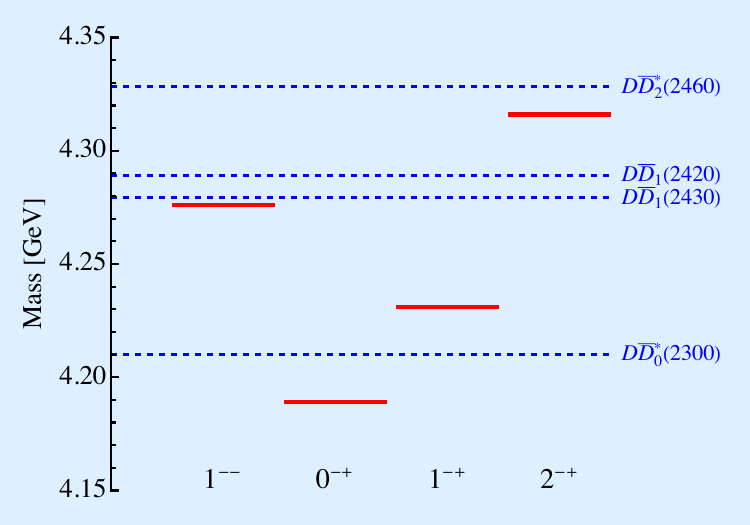}
  \caption{The lowest $c\bar{c}g$ hybrid multiplet spectrum and charmed meson pair mass thresholds.}\label{fig:spectrum}
\end{figure}

The parameters used in the calculation are listed in Table~\ref{tab:parameters}. They are standard values in the literature for heavy quark systems, chosen to reproduce the known charmonium spectrum~\cite{Barnes:1995hc}. The resulting masses for the lowest hybrid multiplet are
\begin{align*}
    m_{0^{-+}} &= 4.189\;\text{GeV},\quad m_{1^{-+}} = 4.231\;\text{GeV},\\
    m_{1^{--}} &= 4.276\;\text{GeV},\quad m_{2^{-+}} = 4.316\;\text{GeV}.
\end{align*}
The spectrum is shown in Fig.~\ref{fig:spectrum} alongside relevant open-charmed meson-pair thresholds.

\begin{table}
    \caption{Parameters used in the spectrum calculation}\label{tab:parameters}
    \begin{ruledtabular}
    \begin{tabular}{cccccc}
        $m_c$ & $m_g$ & $\alpha_s$ & $b$ & $c_H$ & $\sigma$ \\
        \hline
        1.52\;GeV  & 1.05\;GeV  & 0.28     & 0.132\;GeV$^2$ & $-0.58$\;GeV & 1.30\;GeV
    \end{tabular}
    \end{ruledtabular}
\end{table}

\subsection{Comparison with other approaches}

In Table~\ref{tab:mass_comparison}, we compare our mass predictions for the lowest charmonium hybrid multiplet with results from other theoretical approaches, including the bag model, flux tube model, QCD string model, Born-Oppenheimer effective field theory (BOEFT), QCD sum rule, Constituent gluon model, and lattice QCD.

\begin{table}[h]
    \caption{Comparison of charmonium hybrid mass predictions (in GeV) from different theoretical approaches.}\label{tab:mass_comparison}
    \begin{ruledtabular}
        \begin{tabular}{cccccc}
            Approach & $0^{-+}$ & $1^{-+}$ & $2^{-+}$ & $1^{--}$ \\
            \hline
            Bag model~\cite{Barnes:1982tx,Chanowitz:1982qj} & \multicolumn{4}{c}{$\sim 4$} \\
            Flux tube model~\cite{Barnes:1995hc} & \multicolumn{4}{c}{$ 4.1-4.2$} \\
            QCD string model~\cite{Kalashnikova:2016bta} & 4.296 & 4.358 & 4.484 & 4.430 \\
            BOEFT~\cite{Oncala:2017hop} & \multicolumn{4}{c}{4.011} \\
            Constituent model~\cite{Farina:2020slb} & \multicolumn{4}{c}{$\sim 4.4$} \\
            QCD sum rule~\cite{Chen:2013zia} & 3.61 & 3.70 & 4.04 & 3.36 \\
            LQCD ($m_\pi\approx 400$ MeV)~\cite{HadronSpectrum:2012gic} & 4.127 & 4.217 & 4.334 & 4.285 \\
            LQCD ($m_\pi\approx 240$ MeV)~\cite{Cheung:2016bym} & 4.279 & 4.310 & 4.456 & 4.411 \\

            This work & 4.189 & 4.231 & 4.316 & 4.276 \\
        \end{tabular}
    \end{ruledtabular}
\end{table}

In bag model~\cite{Barnes:1982tx,Chanowitz:1982qj}, the lowest energy mode of gluon is TE mode, which can give as high as 1 GeV additional energy, leading to a mass of $c\bar{c}g$ around 4 GeV. The $\mathcal{O}(\alpha_s)$ mass splitting yields the mass ordering as our calculation:
\begin{eqnarray}
    m_{0^{-+}} < m_{1^{-+}} < m_{1^{--}} < m_{2^{-+}}.
\end{eqnarray}

In the flux tube model~\cite{Isgur:1984bm,Barnes:1995hc}, the gluon excitation is modeled as phonon of transverse flux-tube vibration excited in the lowest mode, which may have left- or right-hand polarization. Consequently, the lowest hybrid multiplet contain eight states with quantum number $J^{PC}=1^{\pm\pm}, 0^{\pm\mp}, 1^{\pm\mp}, 2^{\pm\mp}$. In contrast, the constituent gluon model adopted in this work yields only four states with the quantum numbers listed in Table.~\ref{tab:hybrid_states}. The flux tube model predicts the lowest hybrid multiplet to lie around 4.1-4.2 GeV and the spin splitting was estimated to be small~\cite{Merlin:1986tz}.

In Refs.~\cite{Kalashnikova:2008qr,Kalashnikova:2016bta}, the QCD string model was employed to calculate the masses and splittings of the lowest charmonium hybrid states with a TE gluon. The predicted masses, ranging from 4.3 to 4.5 GeV, are higher than our results. The mass splitting pattern is the same as our results and bag model.

The hybrid spectrum was also calculated in the Born-Oppenheimer effective field theory (BOEFT) framework~\cite{Braaten:2014qka,Berwein:2015vca,Oncala:2017hop,Brambilla:2018pyn,Brambilla:2019jfi,Soto:2023lbh}, which makes use of the Born-Oppenheimer potentials obtained from lattice QCD. Here, we refer to the results of Ref.~\cite{Oncala:2017hop} for a comparison, where the lowest hybrid multiplet is predicted to lie around 4.0 GeV, lower than our results and lattice QCD.

Lattice QCD predictions for hybrid masses vary with the pion mass. For $m_\pi\approx 400$ MeV~\cite{HadronSpectrum:2012gic}, the results align well with our predictions, whereas for $m_\pi\approx 240$ MeV~\cite{Cheung:2016bym}, the masses are higher by approximately 100 MeV. In both cases, the mass splitting pattern is consistent with our results.

Although the constituent gluon model was proposed long ago~\cite{Horn:1977rq,LeYaouanc:1984gh}, it remains under development~\cite{Swanson:1998kx,Szczepaniak:2006nx,Guo:2008yz}, and predictions for the $c\bar{c}g$ spectrum remain scarce. We compare our results with the recent work of Ref.~\cite{Farina:2020slb}, which is guided by the QCD Hamiltonian in Coulomb gauge. The predicted spin-averaged mass of the lowest hybrid multiplet is approximately 4.4 GeV, very close to the lattice QCD results with $m_\pi\approx 240$ MeV.

There also exist many QCD sum rule calculations for charmonium hybrids~\cite{Govaerts:1984hc,Govaerts:1985fx,Govaerts:1986pp,Qiao:2010zh,Harnett:2012gs,Chen:2013zia,Wang:2025ypo,Agaev:2025llz}. Here, we refer to Ref.~\cite{Chen:2013zia} for comparison, where prediction for complete lowest hybrid multiplet is given. The predicted masses are significantly lower than our results and lattice QCD.

\section{Open charm decay of charmonium hybrid}

\subsection{Decay mechanism}
The decay of hybrid mesons is governed by the QCD coupling between quarks and gluons. To leading order in perturbation theory, the interaction Hamiltonian in the interaction picture is
\begin{equation}
    H_I = g \int d^4x\, \bar{\psi}(x)\gamma^\mu \frac{\lambda_a}{2} \psi(x) A_\mu^a(x),\label{eq:HI}
\end{equation}
where $g$ is the strong coupling constant, $\lambda_a$ are the Gell-Mann matrices, and $\psi$ and $A_\mu^a$ represent the quark and gluon fields, respectively.

Expanding the field operators in terms of creation and annihilation operators yields the non-relativistic reduction
\begin{align}
    H_I &= g\sum_{s,s',\lambda} \int \frac{d^3p\,d^3p'\,d^3k}{\sqrt{2\omega_k}(2\pi)^{9/2}} (2\pi)^4\delta^{(4)}(p+p'-k) \nonumber\\
    &\quad \times \frac{1}{\sqrt{2E_p}\sqrt{2E_{p'}}} \bar{u}_{\boldsymbol{p}}^s\gamma_\mu \frac{\lambda^a}{2} v_{\boldsymbol{p}'}^{s'} \,
    b^\dagger_{\boldsymbol{p},s} d^\dagger_{\boldsymbol{p}',s'} a_{\boldsymbol{k},a,\lambda} \epsilon^\mu(\lambda,\boldsymbol{k}),
\end{align}
where the single-particle states are normalized as $\langle \boldsymbol{p},s|\boldsymbol{p}',s'\rangle = \delta^{(3)}(\boldsymbol{p}-\boldsymbol{p}')\delta_{ss'}$. In the non-relativistic limit, the Dirac spinor matrix element simplifies to
\begin{equation}
    \frac{1}{\sqrt{2E_p}\sqrt{2E_{p'}}}\bar{u}_{\boldsymbol{p}}^s\gamma_\mu v_{\boldsymbol{p}'}^{s'} \epsilon^\mu(\lambda,\boldsymbol{k}) \approx -\chi_s^\dagger \boldsymbol{\sigma} \tilde{\chi}_{s'} \cdot \boldsymbol{\epsilon}(\lambda,\boldsymbol{k}),
\end{equation}
where $\chi_s$ is the quark Pauli spinor and $\tilde{\chi}_{s'} = -i\sigma_2 \chi_{s'}^*$ represents the antiquark spinor.

A hybrid meson state with definite angular momentum is constructed by coupling the gluon angular wave function from Eq.~\eqref{eq:gluon_angular} with the quark-antiquark spatial and spin wave functions. The resulting state is
\begin{align}
    &|JM[LSlj_g\xi]\rangle  \nonumber\\
    &= \sum_{\substack{i,j,A,\mu,\lambda\\ m_l,m_g,m,\bar{m}\\M_L,M_S}}\frac{1}{2}T_{ij}^A \int d^3p_1\,d^3p_2\,d^3p_g \,\delta^{(3)}(\boldsymbol{p}_1+\boldsymbol{p}_2+\boldsymbol{p}_g) \Psi_{j_g,lm_l}(\boldsymbol{q},k) \nonumber\\
    &\quad \times \sqrt{\frac{2j_g+1}{4\pi}} D_{m_g,\mu}^{j_g*}(\hat{\boldsymbol{k}}) \chi^{(\xi)}_{\mu,\lambda} \langle \tfrac{1}{2}m;\tfrac{1}{2}\bar{m}|SM_S\rangle \nonumber\\
    &\quad \times \langle l m_l;j_g m_g|L M_L\rangle \langle S M_S;L M_L| JM\rangle \nonumber\\
    &\quad \times b^\dagger_{\boldsymbol{p}_1,i,m} d^\dagger_{\boldsymbol{p}_2,j,\bar{m}} a_{\boldsymbol{p}_g,A,\lambda}^\dagger |0\rangle,
\end{align}
where $\frac{1}{2}T^A_{ij}$ is the color wave function. 
The Jacobi momentum are
\begin{eqnarray}
    \boldsymbol{q} = \frac{m_{\bar{c}} \boldsymbol{p}_1 - m_c \boldsymbol{p}_2}{m_c + m_{\bar{c}}} \,,\boldsymbol{k} = \frac{(m_c+m_{\bar{c}})\boldsymbol{p}_g-m_g(\boldsymbol{p}_1+\boldsymbol{p}_2)}{m_c+m_{\bar{c}}+m_g} \,.
\end{eqnarray}
Note that the wave function $\Psi_{j_g,lm_l}(\boldsymbol{q},k)$ depends only on the magnitude of the gluon momentum $\boldsymbol{k}$ since its angular part $Y_{l_g,m_{l_g}}(\hat{\boldsymbol{k}})$ is already encoded in the Wigner $D$-function (see Eq.~\eqref{eq:apY} and Eq.~\eqref{eq:apY2}). The normalization condition $\langle JM[LSlj_g\xi]|JM[LSlj_g\xi]\rangle = \delta^{(3)}(0)$ requires
\begin{equation}\label{eq:Psi_norm}
    \int d^3q\,dk\,k^2 \, |\Psi_{j_g,lm_l}(\boldsymbol{q},k)|^2 = 1.
\end{equation}

A conventional meson state with total momentum $\boldsymbol{P}$ and total angular momentum $J_B$, with $z$-component $M_{J_B}$, is constructed analogously:
\begin{align}
    &|\boldsymbol{P},J_B M_{J_B}\rangle \nonumber\\
    &= \sum_{M_{L_B},M_{S_B},m,m',i}\frac{1}{\sqrt{3}}\int d^3p_1\,d^3p_2 \, \delta^{(3)}(\boldsymbol{p}_1+\boldsymbol{p}_2 - \boldsymbol{P}) \phi_{L_B,M_{L_B}}(\boldsymbol{p}) \nonumber\\
    &\quad \times \langle \tfrac{1}{2}m;\tfrac{1}{2}m'|S_B M_{S_B}\rangle \nonumber\\
    &\quad \times \langle S_B M_{S_B};L_B M_{L_B}|J_B M_{J_B}\rangle b_{\boldsymbol{p}_1,i,m}^\dagger d_{\boldsymbol{p}_2,i,m'}^\dagger |0\rangle,
\end{align}
where $\boldsymbol{p}$ is the relative momentum between the quarks, and the spatial wave function is normalized as $\int d^3p\, |\phi(\boldsymbol{p})|^2 = 1$.

The matrix element for the decay $H \to B + C$ is defined by
\begin{equation}
    \langle B(\boldsymbol{P})C(-\boldsymbol{P})|H_I|H\rangle = (2\pi)\delta^{(4)}(P_H-P_B-P_C)\,\mathcal{M}^{M_{J},M_{J_B},M_{J_C}}(\boldsymbol{P}),
\end{equation}
where the reduced amplitude takes the form
\begin{align}
    &\mathcal{M}^{MM_{J_B}M_{J_C}}(\boldsymbol{P}) \nonumber\\
    &= -\frac{2}{3} \frac{g}{(2\pi)^{3/2}} \int d^3k\,d^3q\,\Psi_{j_g,lm_l}(\boldsymbol{q},k)\sqrt{\frac{2j_g+1}{4\pi}} D^{j_g*}_{m_g,\mu}(\hat{\boldsymbol{k}})\chi_{\mu,\lambda}^{(\xi)} \nonumber\\
    &\quad \times \langle \tfrac{1}{2}m;\tfrac{1}{2}\bar{m}|SM_S\rangle \langle lm_l;j_gm_g|LM_L\rangle \langle SM_S;LM_L|JM\rangle \nonumber\\
    &\quad \times \frac{1}{\sqrt{2\omega_k}} \chi_s^\dagger \boldsymbol{\sigma} \tilde{\chi}_{s'}\cdot\boldsymbol{\epsilon}(\lambda,\boldsymbol{k}) \nonumber\\
    &\quad \times \phi_{L_B M_{L_B}}^*\left(\boldsymbol{q}-\frac{\boldsymbol{k}}{2}-\frac{m_c}{m_q + m_c}\boldsymbol{P}\right) \langle \tfrac{1}{2}m; \tfrac{1}{2}s'|S_B M_{S_B}\rangle \nonumber\\
    &\quad \times \langle S_B M_{S_B};L_BM_{L_B}|J_B M_{J_B}\rangle \nonumber\\
    &\quad \times \phi_{L_CM_{L_C}}^*\left(\boldsymbol{q}+\frac{\boldsymbol{k}}{2}-\frac{m_c}{m_q+m_c}\boldsymbol{P}\right) \langle \tfrac{1}{2}s; \tfrac{1}{2}\bar{m}|S_CM_{S_C}\rangle \nonumber\\
    &\quad \times \langle S_C M_{S_C};L_C M_{L_C}|J_C M_{J_C}\rangle.\label{eq:decay_amplitude}
\end{align}
The summation over repeated indices is implied.
The prefactor $2/3$ arises from the color wave function overlap, and $g/(2\pi)^{3/2}$ follows from our normalization conventions (see Appendix~\ref{app:field_convention}).

The spin matrix element can be expressed in terms of a Wigner $D$-function (see Appendix~\ref{app:sigma}):
\begin{equation}\label{eq:sigma}
    \chi_s^\dagger \boldsymbol{\sigma}\tilde{\chi}_{s'}\cdot \boldsymbol{\epsilon}(\lambda,\boldsymbol{k}) = -\sqrt{2}^{\,|s+s'|}\,D^1_{s+s',\lambda}(\hat{\boldsymbol{k}}).
\end{equation}
For a $j_g=1$ TE gluon ($\xi=-1$), combining this with the gluon angular wave function yields
\begin{align}
    &\sum_{\mu,\lambda}\sqrt{\frac{2j_g+1}{4\pi}} D^{j_g*}_{m_g,\mu}(\hat{\boldsymbol{k}}) \chi_{\mu,\lambda}^{(\xi)}\,  \chi_s^\dagger \boldsymbol{\sigma}\tilde{\chi}_{s'}\cdot \boldsymbol{\epsilon}(\lambda,\boldsymbol{k}) \nonumber\\
    &= -\sqrt{2}^{\,|s+s'|}(-1)^{s+s'} \langle 1 m_g;1 -(s+s')|1 m_{l_g}\rangle Y_{1,m_{l_g}}(\hat{\boldsymbol{k}}),
\end{align}
where $Y_{1,m_{l_g}}(\hat{\boldsymbol{k}})$ represents the angular distribution of the constituent gluon.

The integral over the direction $\hat{\boldsymbol{k}}$ in Eq.~\eqref{eq:decay_amplitude} involves the factor $\phi_B^*(\boldsymbol{k},\boldsymbol{P})\phi_C^*(\boldsymbol{k},\boldsymbol{P}) Y_{1,m_{l_g}}(\hat{\boldsymbol{k}})$. If the two final-state mesons $B$ and $C$ have identical spatial wave functions, the product $\phi_B^*\phi_C^*$ is even under $\boldsymbol{k}\to -\boldsymbol{k}$, while $Y_{1,m_{l_g}}(\hat{\boldsymbol{k}})$ is odd. The angular integral therefore vanishes. This leads to a key selection rule: \textit{a $j_g=1$ TE hybrid cannot decay into two mesons with identical spatial wave functions}~\cite{Farina:2020slb}. This selection rule, also found in other hybrid decay analyses~\cite{Page:1996rj}, has important phenomenological consequences for charmonium hybrid decays.
The decay width is obtained directly from the amplitude in Eq.~\eqref{eq:decay_amplitude} as (see Appendix~\ref{app:decay_width})
\begin{equation}
    \Gamma(H\to BC) = 8\pi^2 \frac{E_B E_C}{M_H} \frac{P}{2J+1} \sum_{M,M_{J_B},M_{J_C}} |\mathcal{M}^{M M_{J_B} M_{J_C}}|^2\,,
\end{equation}
where $P = |\boldsymbol{P}|$ denotes the magnitude of the final-state momentum in the rest frame of the hybrid.

\subsection{Numerical results}
The selection rule derived in the previous subsection has profound implications for the decay patterns of charmonium hybrids with a $j_g=1$ TE gluon. As established, decays to two mesons with identical spatial wave functions are forbidden. Consequently, the $D\bar{D}$ and $D^*\bar{D}^*$ channels are strictly prohibited. Furthermore, since the $D$ and $D^*$ mesons are both $S$-wave $c\bar{q}$ states with very similar spatial wave functions, the decay to $D\bar{D}^*$ is strongly suppressed, though not absolutely forbidden. Therefore, the dominant decay modes for these hybrids, when kinematically allowed, are expected to be into an $S$-wave plus a $P$-wave meson pair (e.g., $D\bar{D}_0^*$, $D\bar{D}_1$, $D\bar{D}_2^*$). If the hybrid mass lies below all $S+P$ thresholds, the state should be narrow.

For the numerical calculation of decay widths, we require explicit forms for the hybrid and meson wave functions $\Psi_{j_g,lm_l}$. In the calculation of hybrid spectrum, we obtain the hybrid ground state wave function in Eq.~\eqref{eq:SHO} in position space with $S$-wave $\lambda$-mode and $S$-wave $\rho$-mode, though the angular wave function $Y_{1m_{l_g}}$ of the gluon is implicit by the TE-mode gluon with quantum number $j_g^{PC}=1^{+-}$. In the calculation of decays, we restore the gluon orbital angular momentum by replacing the $\lambda$-mode SHO wave function with a $P$-wave one, while keeping the $\rho$-mode in the ground state. Thus, the hybrid spatial wave function for the lowest multiplet is of the form:
\begin{equation*}
    \Psi_{j_g,lm_l}(\boldsymbol{q},k) Y_{1,m_{l_g}}(\hat{\boldsymbol{k}}) = \frac{\sqrt{8/3}\,k}{\pi^{3/4}\beta_\rho^{3/2}\pi^{1/4}\beta_\lambda^{5/2}} Y_{1,m_{l_g}}(\hat{\boldsymbol{k}})\, e^{-q^2/2\beta_\rho^2 - k^2/2\beta_\lambda^2},
\end{equation*}
where $\beta_\rho$ and $\beta_\lambda$ are the oscillator parameters obtained from the spectrum calculation.

The conventional meson wave functions are approximated by simple harmonic oscillator (SHO) wave functions:
\begin{equation*}
    R_{n,L}^{\text{SHO}}(p) = \frac{(-1)^n(-i)^L}{\beta^{3/2}}\sqrt{\frac{2n!}{\Gamma\left(n+L+\frac{3}{2}\right)}}\left(\frac{p}{\beta}\right)^L e^{-p^2/2\beta^2}L_n^{L+\frac{1}{2}}\!\left(\frac{p^2}{\beta^2}\right),
\end{equation*}
where $n$ and $L$ are the radial and orbital angular momentum quantum numbers, respectively, and $L_n^{L+\frac{1}{2}}$ denotes an associated Laguerre polynomial.

The oscillator parameters for the charmed mesons and the hybrid are summarized in Table~\ref{tab:beta_params}. The values for $\beta_\rho$ and $\beta_\lambda$ are taken directly from the mass spectrum calculation, while the meson $\beta$ parameters are determined by solving the Salpeter equation~\cite{Chen:2025pvk}.

\begin{table}[h]
    \caption{Oscillator parameters used in the decay width calculation.}
    \label{tab:beta_params}
    \begin{ruledtabular}
        \begin{tabular}{c|ccccc}
            Parameter & $\beta_{D}$ & $\beta_{D^*}$ & $\beta_{D(1P)}$ & $\beta_\rho$ & $\beta_\lambda$ \\
            \hline
            Value (GeV) & 0.574 & 0.496 & 0.385 & 0.432 & 0.625
        \end{tabular}
    \end{ruledtabular}
\end{table}

With the wave functions fixed, the only remaining free parameter is the strong coupling constant $g$, related to the running coupling by $\alpha_S = g^2/(4\pi)$. Given that the constituent gluon mass is approximately 1 GeV, we adopt $\alpha_S = 0.5$ as a typical value of the running coupling at this energy scale.

\begin{figure}[htbp]
    \centering
    \includegraphics[width=0.47\textwidth]{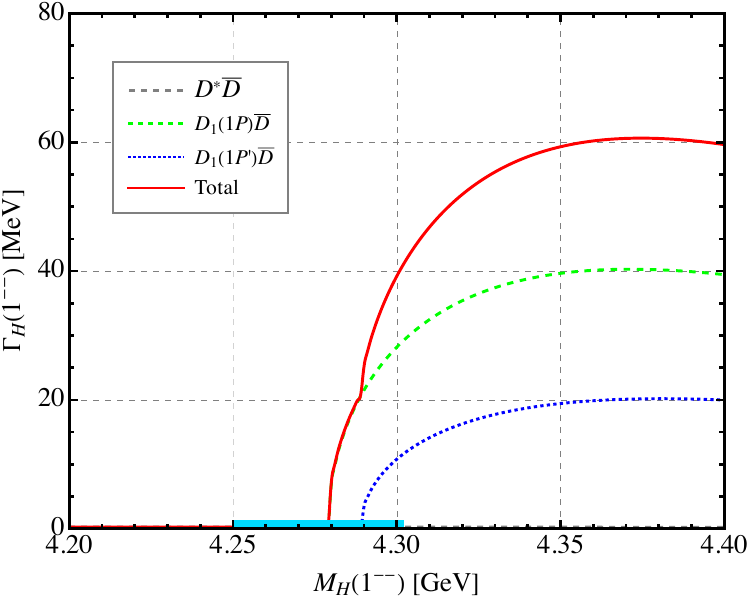}
    \caption{Decay width of the $1^{--}$ hybrid as a function of its mass. The shaded band on the mass axis indicates $\pm25$ MeV around our nominal value for the $1^{--}$ hybrid mass. The inclusion of charge-conjugate channels is implicit.}
    \label{fig:width1}
\end{figure}

\begin{figure}[htbp]
    \centering
    \includegraphics[width=0.47\textwidth]{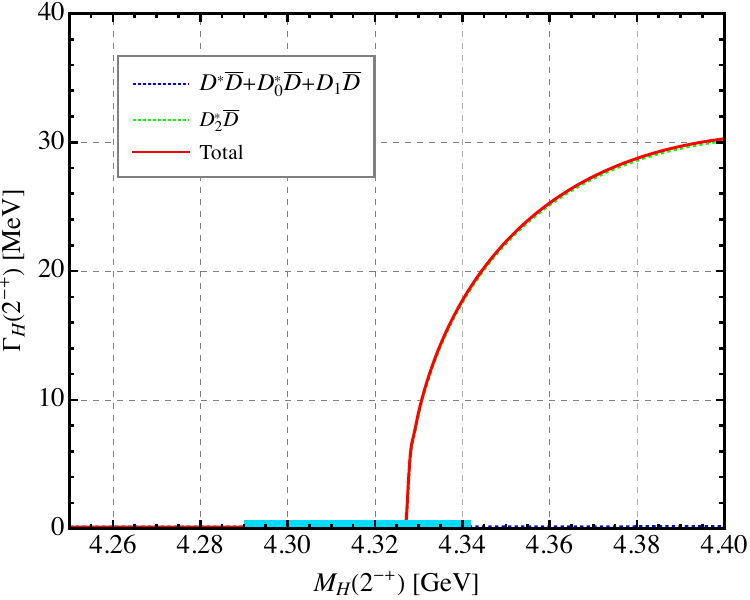}
    \caption{Decay width of the $2^{-+}$ hybrid as a function of its mass. The shaded band on the mass axis indicates $\pm25$ MeV around our nominal value for the $2^{-+}$ hybrid mass. The inclusion of charge-conjugate channels is implicit.}
    \label{fig:width2}
\end{figure}

The two-body open charm decay widths of the $1^{--}$ and $2^{-+}$ hybrid states are shown in Fig.~\ref{fig:width1} and Fig.~\ref{fig:width2}, respectively. The corresponding widths for the $0^{-+}$ and $1^{-+}$ states are very small (less than 1 MeV) and are therefore not displayed.

It should be noted that there exist other decay channels besides the two-body open charm decay, such as hidden-charm decays $H\to (c\bar{c})(gg)\to c\bar{c} + \text{light hadrons}$ and annihilation decays $H\to ng\to \text{light hadrons}$~\cite{Close:2003mb}. Though not estimated explicitly, these decay channels are expected to be suppressed relative to the two-body open charm decay, provided the latter is not suppressed by the selection rule. In what follows, we summarize the main decay properties of the lowest hybrid multiplet based on their two-body open charm decays, with reference to the spectrum presented in Fig.~\ref{fig:spectrum}.
\begin{itemize}
    \item \textbf{$0^{-+}$ hybrid (4.189 GeV):} This state lies below all $S+P$ charmed meson thresholds. With no open decay channels satisfying the selection rules, its total width is expected to be very small. This makes the $0^{-+}$ hybrid an excellent candidate for a narrow resonance.

    \item \textbf{$1^{-+}$ hybrid (4.231 GeV):} Despite lying above the $D\bar{D}_0^*(2300)$ threshold, this decay channel is forbidden by the exotic quantum numbers. The $D\bar{D}^*$ channel is strongly suppressed by the spatial wave function overlap argument. Consequently, the $1^{-+}$ charmonium hybrid is also predicted to be a narrow state, providing a clear experimental signature due to its exotic $J^{PC}$.

    \item \textbf{$1^{--}$ hybrid (4.276 GeV):} Decay to $D\bar{D}_0^*(2300)$ is forbidden by quantum numbers, and $D\bar{D}^*$ is suppressed. The predicted mass lies very close to the $D\bar{D}_1$ threshold, making the width extremely sensitive to the exact mass value. If $M_H > M_D + M_{D_1}$, the $D\bar{D}_1$ channel opens and becomes the dominant decay mode. Fig.~\ref{fig:width1} shows the decay width as a function of mass, with the shaded band indicating $\pm25$ MeV theoretical uncertainty around our nominal value. The width varies from near zero below threshold to several tens of MeV above threshold.

    \item \textbf{$2^{-+}$ hybrid (4.316 GeV):} Decays to $D\bar{D}_0^*$ and $D\bar{D}_1$ proceed via $D$-waves, which strongly suppress these partial widths. The predicted mass is slightly below the $D\bar{D}_2^*$ threshold, so the mass dependence is again crucial. Fig.~\ref{fig:width2} shows the decay width to $D\bar{D}_2^*$ as a function of mass. Once above threshold, this channel dominates, with contributions from $D\bar{D}_0^*$ and $D\bar{D}_1$ being negligible in comparison.
\end{itemize}

These open charm decay patterns describe the main features of the lowest charmonium hybrid multiplet.
The $0^{-+}$ and $1^{-+}$ hybrids should appear as narrow resonances. The $1^{--}$ and $2^{-+}$ states exhibit strong threshold effects, making them promising targets for scanning experiments at facilities such as BESIII, Belle II, and the Super Tau-Charm Facility. Precise measurement of their masses relative to the $D\bar{D}_1$ and $D\bar{D}_2^*$ thresholds would provide tests of the hybrid interpretation. In the next section, we will give more suggestions on how to search for these states experimentally.

There are not many calculations concerning hybrid decay in the literature. Explicit calculations of the open charm decay widths of charmonium hybrids have been performed within the flux tube model~\cite{Page:1998gz} and the constituent gluon model~\cite{Farina:2020slb}. The general decay pattern is consistent with our results: $S+S$ wave charmed meson pair channels are strongly suppressed due to the selection rule. The allowed $S+P$ wave charmed meson pair channels dominate when they are kinematically accessible. Within the flux tube model, the partial widths for the $H(1^{--})\to D_1\bar{D}$ and $H(2^{-+})\to D_2^*\bar{D}$ channels are predicted to be 17.3 MeV and 9 MeV, respectively, for a hybrid mass of 4.4 GeV~\cite{Page:1998gz}. These values are slightly smaller than our results. In the constituent gluon model calculation of Ref.~\cite{Farina:2020slb}, the corresponding widths for $H(1^{--})\to D_1\bar{D}$ and $H(2^{-+})\to D_2^*\bar{D}$ are approximately 41 MeV and 20 MeV at a hybrid mass of about 4.4 GeV, which is consistent with our findings.

\section{More suggestion for searching for the discussed states}
\subsection{The $1^{--}$ hybrid state}

The $1^{--}$ quantum numbers coincide with those of the photon, making $e^+e^-$ annihilation an ideal production mechanism for this hybrid state. However, its coupling to the virtual photon is expected to be significantly smaller than that of conventional $S$-wave charmonium states because the hybrid's $c\bar{c}$ pair is in a spin-singlet configuration ($S_{c\bar{c}}=0$), whereas the virtual photon couples preferentially to spin-triplet configurations. This coupling may nevertheless be larger than that of $D$-wave charmonium states. Given this suppressed direct production, distinguishing the hybrid from conventional charmonium relies critically on their distinct decay patterns.

Our calculation places the $1^{--}$ hybrid mass at approximately 4.27 GeV, with a strong sensitivity of its decay width to the precise mass value. The state couples strongly to the $D\bar{D}_1$ channel, with a width reaching $\sim 40$ MeV near 4.3 GeV. If the mass lies below the $D\bar{D}_1$ threshold, two classes of decay mechanisms remain available: (i) hidden-charm decays $H(1^{--})\to (c\bar{c})(gg)\to c\bar{c} + \text{light hadrons}$, and (ii) annihilation decays $H(1^{--})\to 3g\to \text{light hadrons}$~\cite{Close:2003mb}. However, given the proximity of the predicted mass to the $D\bar{D}_1$ threshold and the strong coupling to this channel, the decay may proceed via an off-shell $D\bar{D}_1$ intermediate state:
\begin{equation*}
    H(1^{--})\to D\bar{D}_1 \to D\bar{D}^*\pi,
\end{equation*}
which could become the dominant mode even when the $D\bar{D}_1$ threshold is not fully open. In this scenario, the state is expected to be relatively narrow.

The observed $Y(4230)$ resonance has long been considered a candidate for this $1^{--}$ hybrid state~\cite{Zhu:2005hp,Close:2005iz}, and its properties are broadly compatible with our predictions. Evidence supporting this interpretation includes: (i) its mass near 4.23 GeV is close to our calculated value, and (ii) its observed width of approximately 20 MeV and its detection in the $D\bar{D}^*\pi$ channel align with hybrid expectations.

However, recent precise experimental data present doubts to the pure hybrid assignment for $Y(4230)$. First, the $1^{--}$ hybrid has $S_{c\bar{c}}=0$, so heavy quark spin symmetry suppresses hidden-charm decays into final states involving charmonia with $S_{c\bar{c}}=1$. Yet experimental measurements yield~\cite{ParticleDataGroup:2024cfk}
\begin{equation}
    \mathcal{R} = \frac{\mathrm{BR}(Y(4230)\to J/\psi\pi\pi)}{\mathrm{BR}(Y(4230)\to h_c\pi\pi)} \approx 2,
\end{equation}
which clearly contradicts this symmetry expectation. Second, BESIII has recently observed $Y(4230)$ in the $e^+e^-\to D\bar{D}$ channel~\cite{BESIII:2024ths,Wang:2025dur}, contradicting the predicted vanishing width for a TE hybrid decaying into $D\bar{D}$. 
The appearance of $Y(4230)$ in $D\bar{D}$ channel is particularly enlightening, as it directly challenges the selection rule that forbids a $j_g=1$ TE hybrid from decaying into two mesons with identical spatial wave functions. This observation suggests that if $Y(4230)$ has a significant hybrid component, it may not be a pure TE hybrid. Instead, it could be a mixture of different hybrid configurations (e.g., TE and TM gluon states) or a mixture of hybrid and conventional charmonium states. Such mixing could relax the selection rules and allow for the observed decay patterns. The hybrid with TM gluon can decay into $D\bar{D}$, while the $J^{PC}=1^{--}$ hybrid with TM gluon have $l_{c\bar{c}}=1$ and its energy is expected to be relatively far from the TE hybrid. How large the mixing between these two hybrid configurations is an interesting question. The mixing between the hybrid and conventional charmonium states with nearby masses is also possible, which can further complicate the decay patterns. A detailed analysis of such mixing effects would be necessary to fully understand the nature of $Y(4230)$ and its compatibility with the hybrid interpretation.
%These tensions suggest that interpreting $Y(4230)$ as a pure hybrid state remains problematic. The discrepancies might be resolved by including additional mechanisms, such as mixing between the hybrid and conventional charmonium states with nearby masses.

\subsection{The $(0,1,2)^{-+}$ hybrid states}

In the light-flavor sector, several candidates for hybrid mesons with exotic quantum numbers $J^{PC}=1^{-+}$ have been proposed, including $\pi_1(1400)$, $\pi_1(1600)$, $\pi_1(2015)$, and the more recent $\eta_1(1855)$. The $\pi_1$ states have been observed in diffractive $\pi N$ scattering processes~\cite{Aoyagi:1993kn,E852:1997gvf,E852:1998mbq,CrystalBarrel:1998cfz,IHEP-Brussels-LosAlamos-AnnecyLAPP:1988iqi,Khokhlov:2000tk,VES:2001rwn,Baker:2003jh,COMPASS:2009xrl,CLEO:2011upl,E852:2004gpn,E852:2004rfa} and in $N\bar{N}$ annihilation~\cite{CrystalBarrel:1998cfz,CrystalBarrel:1999reg,CrystalBarrel:2019zqh}. Such production mechanisms are not directly applicable to charmonium hybrids. Notably, the $\eta_1(1855)$ was discovered in the radiative annihilation process $J/\psi\to\gamma \eta_1(1855)$~\cite{BESIII:2022iwi,BESIII:2022riz}. This is significant because quarkonium annihilation provides a gluon-rich environment, making it an ideal laboratory for producing hybrid states.

\begin{figure}[htbp]
    \centering
    \includegraphics[width=0.45\textwidth]{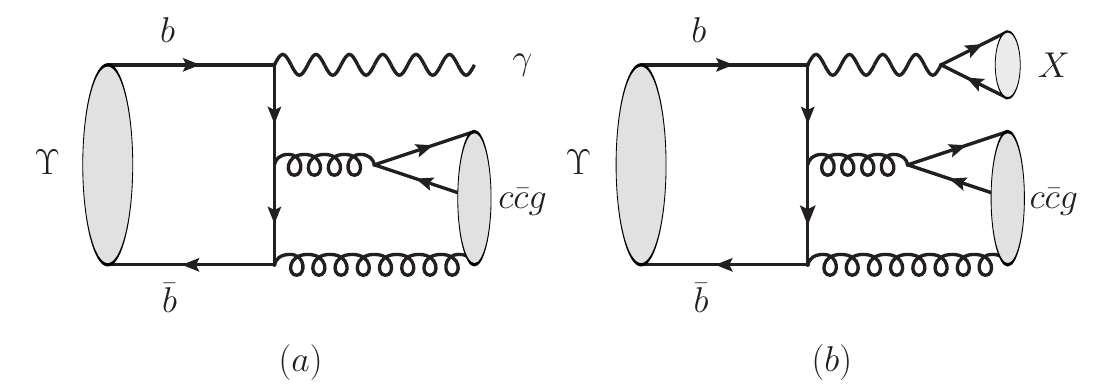}
    \caption{Production of a charmonium hybrid through annihilation decay of $\Upsilon$ states: (a) $\gamma H_{c\bar{c}}$ and (b) $X H_{c\bar{c}}$, where $X$ denotes possible single or multiple hadrons.}
    \label{fig:Upsilon}
\end{figure}

For a $1^{--}$ bottomonium state such as $\Upsilon(nS)$, one of the dominant annihilation processes proceeds via $\gamma gg$ intermediate states (see Fig.~\ref{fig:Upsilon}). The subsequent hadronization of this gluonic system can produce charmonium hybrids. Crucially, the production of conventional charmonium in such processes requires the simultaneous creation of two $c\bar{c}$ pairs, since the $c\bar{c}$ pair produced directly from gluons is in a color-octet configuration and must neutralize its color. This additional suppression makes hybrid production comparatively more favorable in quarkonium annihilation. Annihilation decays of $\Upsilon(nS)$ states,
\begin{equation*}
    \Upsilon(nS)\to\gamma H_{c\bar{c}}\quad\text{and}\quad \Upsilon(nS)\to X H_{c\bar{c}},
\end{equation*}
where $X$ denotes one or more light hadrons, therefore offer a promising pathway to produce hybrid states with quantum numbers $J^{PC}=(0,1,2)^{-+}$. The presence of a photon in the final state selects $C=+1$ hybrids, precisely the $(0,1,2)^{-+}$ multiplet.

The decays of $(0,1,2)^{-+}$ charmonium hybrids may proceed via two mechanisms: hidden-charm decays $H(J^{-+})\to (c\bar{c})(gg)\to c\bar{c}+\text{light hadrons}$, and annihilation decays $H(J^{-+})\to 2g\to \text{light hadrons}$. These modes may have sizable branching fractions, particularly for the $0^{-+}$ and $1^{-+}$ states whose decays to $S+P$ charmed meson pairs are forbidden either by quantum numbers or by phase space.

For hidden-charm decays, the final-state charmonium must have the same $C$-parity as the hybrid because the $gg$ system has even $C$-parity. Heavy-quark spin symmetry further favors $S_{c\bar{c}}=1$ in the transition. These constraints select hidden-charm decays involving $\chi_{cJ}$ states:
\begin{equation*}
    H(J^{-+})\to \chi_{cJ}\pi\pi,\; \chi_{cJ}K\bar{K},\; \ldots
\end{equation*}
For annihilation decays into light hadrons, many final states are possible, and the decay pattern may resemble that of the $\eta_c$, as both proceed via $gg$ annihilation.

The $2^{-+}$ charmonium hybrid is particularly interesting. Our calculated mass lies very close to the $D\bar{D}_2^*(2460)$ threshold. Consequently, the $D\bar{D}_2^*(2460)$ and its subsequent decay $D\bar{D}^{(*)}\pi$ channels are a promising discovery mode if the mass lies above the corresponding thresholds. Moreover, the proximity to this threshold implies that rescattering effects into hidden-charm final states could be significant. Heavy-quark spin symmetry restricts the $c\bar{c}$ pair to $S_{c\bar{c}}=1$, making processes such as
\begin{equation*}
    H(2^{-+})\to D\bar{D}_2^*(2460) \to J/\psi \omega,\; J/\psi\phi
\end{equation*}
promising channels for experimental searches. Therefore, $\Upsilon\to\gamma J/\psi\omega(\phi)$ and $\Upsilon\to X J/\psi\omega(\phi)$ decays provide particularly clean signatures. Future high-luminosity $B$-factories, such as Belle II, are well positioned to explore these decay channels and potentially discover the $2^{-+}$ charmonium hybrid.

\section{Summary}

Charmonium hybrids offer a unique window into the gluonic degrees of freedom in QCD. In this work, we have systematically investigated the spectrum and decay properties of charmonium hybrid mesons within a constituent gluon model, treating the hybrid as a $c\bar{c}g$ three-body system with a TE mode gluon. Our calculations yield masses for the lowest hybrid multiplet: $4.189$ GeV ($0^{-+}$), $4.231$ GeV ($1^{-+}$), $4.276$ GeV ($1^{--}$), and $4.316$ GeV ($2^{-+}$). These states lie in the $4.2$--$4.3$ GeV region, in close proximity to several important charmed meson pair thresholds.

The two-body open-charm decays of these hybrids were calculated using the QCD quark-gluon coupling mechanism. The dominant decay modes are to $S$-wave plus $P$-wave charmed meson pairs, a consequence of the selection rule that forbids decays into two mesons with identical spatial wave functions. The $1^{--}$ hybrid decays predominantly to $D\bar{D}_1$ when kinematically allowed, with a width that exhibits strong mass dependence near threshold. The $2^{-+}$ hybrid favors $D\bar{D}_2^*$ decay when phase space permits. The $0^{-+}$ and $1^{-+}$ hybrids are expected to be narrow states due to phase space limitations and quantum number constraints.

The mass and decay properties of the $Y(4230)$ resonance show reasonable agreement with our predicted $1^{--}$ hybrid. However, recent experimental data on its hidden-charm decays and $D\bar{D}$ production challenge a pure hybrid interpretation. This tension suggests that more complex scenarios, such as mixing between hybrid and conventional charmonium states, may be necessary to fully reconcile theory with observation.

The $(0,1,2)^{-+}$ hybrids present promising discovery opportunities in $\Upsilon$ annihilation decays, specifically through $\Upsilon \to \gamma H_{c\bar{c}}$ and $\Upsilon \to X H_{c\bar{c}}$ channels. Their distinctive decay patterns, combined with heavy-quark spin symmetry constraints, provide powerful tools for distinguishing hybrids from conventional charmonium states. The $0^{-+}$ and $1^{-+}$ hybrids may be observable as narrow resonances in hidden-charm channels such as $\chi_{cJ}\pi\pi$ and $\chi_{cJ}K\bar K$. The $2^{-+}$ hybrid may be accessible through its direct decay to $D\bar{D}_2^*$ and $D\bar{D}^{(*)}\pi$, or via rescattering of the $D\bar{D}_2^*$ channel into hidden-charm final states like $J/\psi \omega$ and $J/\psi \phi$.

In conclusion, charmonium hybrids constitute a unique laboratory for exploring gluonic excitations in QCD. The theoretical predictions presented here, combined with ongoing and future experimental efforts at BESIII, Belle II, and next-generation facilities, offer promising prospects for the discovery and characterization of these exotic hadrons. A combined analysis of multiple decay channels and production mechanisms will be essential to establish the hybrid nature of any observed candidate.

\begin{acknowledgements}
This work is supported by the National Natural Science Foundation of China under Grants No.~12335001 and No.~12247101, the 111 Center under Grant No.~B20063, the Natural Science Foundation of Gansu Province (No.~22JR5RA389, No.~25JRRA799), the Talent Scientific Fund of Lanzhou University, the fundamental Research Funds for the Central Universities (No.~lzujbky-2023-stlt01), the project for top-notch innovative talents of Gansu province and Lanzhou City High-Level Talent Funding.
\end{acknowledgements}

\appendix

\section{The TE and TM gluons}\label{app:TEM_gluon}
The gluon spin couples to its orbital angular momentum $l_g$ to yield the total gluon angular momentum $j_g$. The gluon wave function is an eigenstate of $\hat{J}^2_g$, $\hat{J}_{gz}$, $\hat{L}_g^2$ and $\hat{S}_g^2$, and can be expressed in terms of the vector spherical harmonics
\begin{equation}\label{eq:apY}
    \boldsymbol{Y}_{j_g,m_g}^{l_g}(\theta,\phi) \equiv \sum_{m_{l_g},\sigma} \langle l_g m_{l_g};1 \sigma |j_g m_g\rangle Y_{l_g,m_{l_g}}(\theta,\phi) \boldsymbol{\epsilon}(\sigma) \,,
\end{equation}
where $\boldsymbol{\epsilon}(\sigma)$ is the gluon polarization function. The transverse and longitudinal components can be separated as follows~\cite{doi:10.1142/0270}: 
\begin{eqnarray}\label{eq:YTEM}
    \boldsymbol{Y}^{(\text{TE})}_{j_g,m_g} &=& \boldsymbol{Y}^{j_g}_{j_g,m_g} \,, \nonumber\\
    \boldsymbol{Y}^{(\text{TM})}_{j_g,m_g} &=& \sqrt{\frac{j_g+1}{2j_g+1}}\boldsymbol{Y}^{j_g-1}_{j_g,m_g} + \sqrt{\frac{j_g}{2j_g+1}}\boldsymbol{Y}^{j_g+1}_{j_g,m_g} \,, \nonumber\\
    \boldsymbol{Y}^{(\text{long})}_{j_g,m_g} &=& \sqrt{\frac{j_g}{2j_g+1}}\boldsymbol{Y}^{j_g-1}_{j_g,m_g} - \sqrt{\frac{j_g+1}{2j_g+1}}\boldsymbol{Y}^{j_g+1}_{j_g,m_g}\,,\nonumber
\end{eqnarray}
where the TE and TM spherical harmonics satisfy $\hat{\boldsymbol{n}}\cdot \boldsymbol{Y}^{(\text{TE/TM})}_{j_g,m_g}=0$ and the longitudinal spherical harmonics satisfy $\hat{\boldsymbol{n}}\times \boldsymbol{Y}^{(\text{long})}_{j_g,m_g} = 0$, with $\hat{\boldsymbol{n}}$ the unit vector in the $(\theta,\phi)$ direction. These spherical harmonics are mutually orthogonal.
Expressing the gluon polarization in the helicity basis via $\boldsymbol{\epsilon}(\sigma) = \sum_{\lambda}D^{1*}_{\sigma,\lambda}(\phi,\theta,0) \boldsymbol{\epsilon}({\lambda},\hat{\boldsymbol{n}})$ and reducing two Wigner-$D$ matrices to one gives
%\begin{widetext}
\begin{eqnarray}
    \boldsymbol{Y}^{(\text{TE})}_{j_g,m_g}(\theta,\phi) &=& \sqrt{\frac{2j_g+1}{8\pi}} \left( D^{j_g*}_{m_g,-1}(\phi,\theta,0)\boldsymbol{\epsilon}({-1},\hat{\boldsymbol{n}}) \right.\nonumber\\&&\left.- D^{j_g*}_{m_g,1}(\phi,\theta,0)\boldsymbol{\epsilon}({1},\hat{\boldsymbol{n}}) \right)\,, \\
    \boldsymbol{Y}^{(\text{TM})}_{j_g,m_g}(\theta,\phi) &=& \sqrt{\frac{2j_g+1}{8\pi}} \left(  D^{j_g*}_{m_g,1}(\phi,\theta,0)\boldsymbol{\epsilon}({-1},\hat{\boldsymbol{n}})  \right.\nonumber\\&&\left.+ D^{j_g*}_{m_g,-1}(\phi,\theta,0)\boldsymbol{\epsilon}({1},\hat{\boldsymbol{n}}) \right) \,.
\end{eqnarray}
%\end{widetext}
The above expression can be rewritten as
\begin{equation}\label{eq:apY2}
    \boldsymbol{Y}_{j_g,m_g}^{(\xi)}(\theta,\phi) = \sum_{\mu,\lambda}\sqrt{\frac{2j_g+1}{4\pi}} D_{m_g,\mu}^{j_g*}(\phi,\theta,0)\chi_{\mu,\lambda}^{(\xi)} \boldsymbol{\epsilon}(\lambda,\hat{\boldsymbol{n}}) \,,
\end{equation}
where $\xi=-1$ corresponds to a TE gluon and $\xi =1$ corresponds to a TM gluon. The explicit form of $\chi_{\lambda,\mu}^{(\xi)}$ is
\begin{equation}
    \chi_{\lambda,\mu}^{(-)} = -\frac{\lambda}{\sqrt{2}}\delta_{\lambda,\mu},\; \chi_{\lambda,\mu}^{(+)} = \frac{1}{\sqrt{2}}\delta_{\lambda,\mu} \,.
\end{equation}

\section{The field theory convention and decay amplitude}\label{app:field_convention}
In deriving the decay amplitude, we have adopted the following field theory convention. The spinor field has the form
\begin{eqnarray}
    \psi(x) &=& \sum_{s=1}^2 \int \frac{d^3p}{(2\pi)^{3/2}} \frac{1}{\sqrt{2E_{\boldsymbol{p}}}} \left( b_{\boldsymbol{p},s} u_{\boldsymbol{p}}^s  e^{-ipx} + d^{\dagger }_{\boldsymbol{p},s} v_{\boldsymbol{p}}^s e^{ipx} \right)\,, \nonumber\\
    \bar{\psi}(x) &=& \sum_{s=1}^2 \int \frac{d^3p}{(2\pi)^{3/2}} \frac{1}{\sqrt{2E_{\boldsymbol{p}}}} \left( d_{\boldsymbol{p},s} \bar{v}_{\boldsymbol{p}}^s  e^{-ipx} + b^{\dagger }_{\boldsymbol{p},s} \bar{u}_{\boldsymbol{p}}^s e^{ipx} \right) \,,
\end{eqnarray}
where the spinors are given in the Dirac representation by
\begin{eqnarray}
    u(p,s)=\sqrt{E+m} \begin{pmatrix}
        \chi_s \\
        \frac{\boldsymbol{\sigma}\cdot\boldsymbol{p}}{E+m}\chi_s
    \end{pmatrix} ,
    v(p,s) = \sqrt{E+m} \begin{pmatrix}
        \frac{\boldsymbol{\sigma}\cdot\boldsymbol{p}}{E+m} \tilde{\chi}_s \\ \tilde{\chi}_s
    \end{pmatrix}\,.
\end{eqnarray}
where the $\chi_s$ is the quark spinor and $\tilde{\chi}_{s'}=-i\sigma_2\chi_{s'}^{*}$ is the antiquark spinor.
The creation and annihilation operators obey the anticommutation rules
\begin{eqnarray}
    \{b_{\boldsymbol{p},s},b_{\boldsymbol{q},s'}^\dagger\} = \{ d_{\boldsymbol{p},s},d_{\boldsymbol{q},s'}^\dagger \} = \delta^{(3)}(\boldsymbol{p}-\boldsymbol{q})\delta_{s,s'} \,,
\end{eqnarray}
with all other anticommutators equal to zero. Then the equal time anticommutation relation for the $\psi$ and $\psi^\dagger$ are
\begin{eqnarray}
    \{ \psi_a(\boldsymbol{x}),\psi_b^\dagger(\boldsymbol{y}) \} &=& \delta^{(3)}(\boldsymbol{x}-\boldsymbol{y})\delta_{a,b} \,,\nonumber\\
    \{ \psi_a(\boldsymbol{x}),\psi_b(\boldsymbol{y}) \} &=& \{ \psi_a^\dagger(\boldsymbol{x}),\psi_b^\dagger(\boldsymbol{y}) \} = 0  \,.
\end{eqnarray}
The normalization condition for the single particle state created by the creation operators is then
\begin{eqnarray}
    \langle \boldsymbol{p},s| \boldsymbol{q},s'\rangle = \langle 0| \{b_{\boldsymbol{p},s},b^\dagger_{\boldsymbol{q},s'}\} |0\rangle = \delta^{(3)}(\boldsymbol{p}-\boldsymbol{q})\delta_{s,s'} \,.
\end{eqnarray}
Similarly, the gluon field have the form
\begin{eqnarray}
    A_a^\mu(x) &=& \sum_{\lambda=1}^{2} \int \frac{d^3 k}{(2\pi)^{3/2}}\,  \frac{1}{\sqrt{2\omega_k}} \left( a_{\boldsymbol{k},\lambda,a} \epsilon^\mu(\boldsymbol{k},\lambda)e^{-ikx} \right.\nonumber\\
    &&\left. + a_{\boldsymbol{k},\lambda,a}^\dagger \epsilon^{\mu*}(\boldsymbol{k},\lambda)e^{ikx} \right) \,.
\end{eqnarray}
The gluon state created by the creation operator is normalized in the same way as the quark state.

In these conventions, the decay amplitude is related to the matrix element of the interaction Hamiltonian $H_I$ as
\begin{equation}
    \langle BC|H_I|H\rangle = (2\pi)\delta^{(4)}(P_B+P_C-P_H)\mathcal{M}(\boldsymbol{P}) \,.
\end{equation}
Using the interaction Hamiltonian in Eq.~\eqref{eq:HI}, the hybrid and meson states, the decay amplitude can be derived as follows:
\begin{align*}
& (2\pi)\delta^{(4)}(P_B+P_C-P_H)\mathcal{M}(\boldsymbol{P}) \\
&= \Bigg[ \int d^3 p_{B_1} d^3 p_{B_2} d^3 p_{C_1} d^3 p_{C_2}\,  \delta^{(3)}(\boldsymbol{p}_{B_1}+\boldsymbol{p}_{B_2}-\boldsymbol{P}) \nonumber\\&\quad\times\delta^{(3)}(\boldsymbol{p}_{C_1}+\boldsymbol{p}_{C_2}-\boldsymbol{P}_C) \dots \Bigg] \\
&\quad \times \Bigg[ g \int \frac{d^3 p\,d^3p'\,d^3k'}{(2\pi)^{9/2}} (2\pi)^4\delta^{(4)}(p+p'-k')\dots \Bigg] \\
&\quad \times \Bigg[ \int d^3p_1\,d^3p_2\, d^3k \, \delta^{(3)}(\boldsymbol{p}_1+\boldsymbol{p}_2 + \boldsymbol{k})\dots \Bigg] \\
&\quad \times \langle \boldsymbol{p}_{B_1}\boldsymbol{p}_{B_2};\boldsymbol{p}_{C_1}\boldsymbol{p}_{C_2}| b^\dagger_{\boldsymbol{p}} d^\dagger_{\boldsymbol{p}'}a_{\boldsymbol{k}'} |\boldsymbol{p}_1\boldsymbol{p}_2\boldsymbol{k}\rangle \\
&= g \int d^3p_1\,d^3p_2\,d^3k \frac{(2\pi)^4}{(2\pi)^{9/2}}\delta^{(4)}(P_B-p_1+P_C-p_2-k) \nonumber\\&\quad\times\delta^{(3)}(\boldsymbol{p}_1+\boldsymbol{p}_2+\boldsymbol{k})\dots \\
&= (2\pi)\delta^{(4)}(P_B+P_C-P_H)\frac{g}{(2\pi)^{3/2}} \int d^3q\,d^3k \dots
\end{align*}
where the ellipses represent the wave functions and spinor structures in the integrand. The final expression of the decay amplitude is given in Eq.~\eqref{eq:decay_amplitude}.

\section{The derivation of $\chi_s^\dagger \boldsymbol{\sigma}\tilde{\chi}_{s'}\cdot \boldsymbol{\epsilon}(\lambda,\boldsymbol{k})$}\label{app:sigma}

Here, the detailed derivation of Eq.~\eqref{eq:sigma} is given. First, we note that
\begin{eqnarray}
    \chi^\dagger_{\sigma} \boldsymbol{S} \chi_{\sigma'}\cdot \boldsymbol{\epsilon}(\lambda,\boldsymbol{k}) &=& \sum_\mu\chi^\dagger_{\sigma} S_\mu \chi_{\sigma'}  D^1_{\mu,\lambda}(\hat{\boldsymbol{k}}) \nonumber \\
    &=& \sum_\mu\frac{\sqrt{3}}{2} \langle \frac{1}{2}\sigma';1 \mu|\frac{1}{2}\sigma\rangle D_{\mu,\lambda}^1(\hat{\boldsymbol{k}}) \nonumber \\
    &=& -\text{sign}(-\sigma') \frac{1}{2}\sqrt{2}^{|\sigma-\sigma'|} D^1_{\sigma-\sigma',\lambda}(\hat{\boldsymbol{k}}) \,,
\end{eqnarray}
where $S=\boldsymbol{\sigma}/2$. In the meanwhile, we have
\begin{eqnarray}
    \tilde{\chi}_{s'} = -i\sigma_2\chi_{s'}^* = \text{sign}(s')\chi_{-s'}\,.
\end{eqnarray}
Combining the above two equation, we get
\begin{equation}
    \chi_s^\dagger \boldsymbol{\sigma}\tilde{\chi}_{s'}\cdot \boldsymbol{\epsilon}(\lambda,\boldsymbol{k}) = -\sqrt{2}^{|s+s'|} D^1_{s+s',\lambda}(\hat{\boldsymbol{k}})
\end{equation}

\section{The convention of $S$-matrix and decay width}\label{app:decay_width}
The explicit formula of decay width depends on the convention of $S$-matrix and normalization of states. Here we summarize the nonrelativistic convention used in this work. The $S$-matrix is defined as
\begin{equation}
    \langle \boldsymbol{p}_B\boldsymbol{p}_C|S^{NR}|\boldsymbol{p}_A\rangle = \langle \boldsymbol{p}_B\boldsymbol{p}_C|\boldsymbol{p}_A\rangle - 2\pi i\delta^{(4)}\left(p_A- p_B-p_C\right)\mathcal{M}_{A\to BC} \,,
\end{equation}
where the normalization of the single particle state is 
\begin{eqnarray}
    \langle \boldsymbol{p}|\boldsymbol{p}'\rangle = \delta^{(3)}(\boldsymbol{p}-\boldsymbol{p}')\,.
\end{eqnarray}
Note that in the standard relativistic convention~\cite{Peskin:1995ev}, the normalization of single particle state and $S$-matrix are
\begin{eqnarray}
    _R\langle \boldsymbol{p}|\boldsymbol{p}'\rangle_R &=& (2\pi)^3 (2E_{\boldsymbol{p}})\delta^{(3)}(\boldsymbol{p}-\boldsymbol{p}')\,,\nonumber\\
    _R\langle \boldsymbol{p}_B\boldsymbol{p}_C|S^{R}|\boldsymbol{p}_A\rangle_R &=& _R\langle \boldsymbol{p}_B\boldsymbol{p}_C|\boldsymbol{p}_A\rangle_R \nonumber\\
    && + i(2\pi)^4\delta^{(4)}\left(p_A- p_B-p_C\right)T_{A\to BC} \,,
\end{eqnarray}
The two-body decay width is given by
\begin{eqnarray}
    \Gamma_{A\to BC} &=& \frac{P}{8\pi m_A^2}\frac{1}{2J_A+1}\sum_{spin}|T_{A\to BC}|^2 \,,
\end{eqnarray}
where $P$ is the momentum of the final state particles in the rest frame of particle $A$. The $S$-matrix in the two conventions differ by a factor that can be determined by their different normalization condition of particle states.
The relation between the nonrelativistic amplitude $\mathcal{M}_{A\to BC}$ and the relativistic amplitude $T_{A\to BC}$ is
\begin{eqnarray}
    (2\pi)^4T_{A\to BC} &=& -(2\pi)\prod_{i=A,B,C}\left[(2\pi)^{3/2}\sqrt{2E_i}\right]\, \mathcal{M}_{A\to BC} \,.
\end{eqnarray}
Then we obtain the decay width in terms of the nonrelativistic amplitude:
\begin{eqnarray}
    \Gamma_{A\to BC} &=& \frac{8\pi^2}{m_A} E_B E_C\frac{P}{2J_A+1}\sum_{spin}|\mathcal{M}_{A\to BC}|^2 \,.
\end{eqnarray}

\iffalse
The above amplitude can be expressed in terms of the partial wave amplitude as
\begin{eqnarray}
    & & \mathcal{M}^{M_A,M_B,M_C}(\boldsymbol{P}) \nonumber\\
    &=& \sum_{\substack{L,S\\M_S,M_L}} \langle L M_L S M_S|J_A M_A\rangle \langle J_B M_B J_C M_C|S M_S\rangle \nonumber\\
    & &\times Y_{L M_L}(\hat{\boldsymbol{P}}) \mathcal{M}^{LS}(P) \,.
\end{eqnarray}
From the above relation, we can get
\begin{eqnarray}
    \sum_{M_{J_A},M_{J_B},M_{J_C}} |\mathcal{M}^{M_{J_A},M_{J_B},M_{J_C}}(\boldsymbol{P})|^2 &=& \sum_{L,S}  \frac{2J_A+1}{4\pi} |\mathcal{M}^{LS}(P)|^2 \,,\nonumber
\end{eqnarray}
so the decay width in terms of the partial wave amplitude is
\begin{eqnarray}
    \Gamma_{A\to BC} &=& 2\pi \frac{E_B E_C}{m_A} P \sum_{L,S} |\mathcal{M}^{LS}(P)|^2 \,.
\end{eqnarray}
This is the formula used in the main text.
\fi

\bibliography{docbib}

@article{IHEP-Brussels-LosAlamos-AnnecyLAPP:1988iqi,
    author = "Alde, D. and others",
    editor = "Kotthaus, R. and Kuhn, Johann H.",
    collaboration = " IHEP-IISN-LANL-LAPP",
    title = "{Evidence for a $1^{-+}$ Exotic Meson}",
    reportNumber = "CERN-EP-88-15",
    doi = "10.1016/0370-2693(88)91686-3",
    journal = "Phys. Lett. B",
    volume = "205",
    pages = "397",
    year = "1988"
}

@article{Aoyagi:1993kn,
    author = "Aoyagi, H. and others",
    editor = "Nakai, K. and Ohshima, T.",
    title = "{Study of the $\eta \pi^-$ system in the $\pi^- p$ reaction at 6.3 GeV/c}",
    doi = "10.1016/0370-2693(93)90456-R",
    journal = "Phys. Lett. B",
    volume = "314",
    pages = "246--254",
    year = "1993"
}

@article{E852:1997gvf,
    author = "Thompson, D. R. and others",
    collaboration = "E852",
    title = "{Evidence for exotic meson production in the reaction $\pi^- p \to \eta \pi^- p$ at 18 GeV/c}",
    eprint = "hep-ex/9705011",
    archivePrefix = "arXiv",
    doi = "10.1103/PhysRevLett.79.1630",
    journal = "Phys. Rev. Lett.",
    volume = "79",
    pages = "1630--1633",
    year = "1997"
}

@article{VES:2001rwn,
    author = "Dorofeev, Valery and others",
    editor = "Amelin, Dmitry and Zaitsev, Alexander M.",
    collaboration = "VES",
    title = "{The $J^{PC} = 1^{-+}$ hunting season at VES}",
    eprint = "hep-ex/0110075",
    archivePrefix = "arXiv",
    doi = "10.1063/1.1482444",
    journal = "AIP Conf. Proc.",
    volume = "619",
    number = "1",
    pages = "143--154",
    year = "2002"
}

@article{CrystalBarrel:1998cfz,
    author = "Abele, A. and others",
    collaboration = "Crystal Barrel",
    title = "{Exotic $\eta\pi$ state in $\bar{p}d$ annihilation at rest into $\pi^- \pi^0 \eta p$(spectator)}",
    doi = "10.1016/S0370-2693(98)00123-3",
    journal = "Phys. Lett. B",
    volume = "423",
    pages = "175--184",
    year = "1998"
}

@article{CrystalBarrel:1999reg,
    author = "Abele, A. and others",
    collaboration = "Crystal Barrel",
    title = "{Evidence for a $\pi\eta$ P wave in $\bar{p}p$ annihilations at rest into $\pi^0 \pi^0 \eta$}",
    doi = "10.1016/S0370-2693(98)01544-5",
    journal = "Phys. Lett. B",
    volume = "446",
    pages = "349--355",
    year = "1999"
}

@article{CrystalBarrel:2019zqh,
    author = "Albrecht, M. and others",
    collaboration = "Crystal Barrel",
    title = "{Coupled channel analysis of $\bar{p}p \to \pi^0\pi^0\eta$, $\pi^0\eta\eta$ and $K^+K^-\pi^0$ at 900 MeV/c and of $\pi\pi$-scattering data}",
    eprint = "1909.07091",
    archivePrefix = "arXiv",
    primaryClass = "hep-ex",
    doi = "10.1140/epjc/s10052-020-7930-x",
    journal = "Eur. Phys. J. C",
    volume = "80",
    number = "5",
    pages = "453",
    year = "2020"
}

@article{E852:1998mbq,
    author = "Adams, G. S. and others",
    collaboration = "E852",
    title = "{Observation of a new $J^{PC} = 1^{-+}$ exotic state in the reaction $\pi^- p \to \pi^+ \pi^- \pi^- p$ at 18 GeV/c}",
    doi = "10.1103/PhysRevLett.81.5760",
    journal = "Phys. Rev. Lett.",
    volume = "81",
    pages = "5760--5763",
    year = "1998"
}

@article{Khokhlov:2000tk,
    author = "Khokhlov, Yu. A.",
    editor = "Faldt, G. and Hoistad, B. and Kullander, S.",
    collaboration = "VES",
    title = "{Study of $X(1600)$ $1^{-+}$ hybrid}",
    doi = "10.1016/S0375-9474(99)00663-6",
    journal = "Nucl. Phys. A",
    volume = "663",
    pages = "596--599",
    year = "2000"
}

@article{COMPASS:2009xrl,
    author = "Alekseev, M. and others",
    collaboration = "COMPASS",
    title = "{Observation of a $J^{PC} = 1^{-+}$ exotic resonance in diffractive dissociation of 190 GeV/c $\pi^-$ into $\pi^- \pi^- \pi^+$}",
    eprint = "0910.5842",
    archivePrefix = "arXiv",
    primaryClass = "hep-ex",
    reportNumber = "CERN-PH-EP-2009-018",
    doi = "10.1103/PhysRevLett.104.241803",
    journal = "Phys. Rev. Lett.",
    volume = "104",
    pages = "241803",
    year = "2010"
}

@article{E852:2004gpn,
    author = "Kuhn, Joachim and others",
    collaboration = "E852",
    title = "{Exotic meson production in the $f_1(1285) \pi^-$ system observed in the reaction $\pi^- p \to \eta \pi^+ \pi^- \pi^- p$ at 18 GeV/c}",
    eprint = "hep-ex/0401004",
    archivePrefix = "arXiv",
    doi = "10.1016/j.physletb.2004.05.032",
    journal = "Phys. Lett. B",
    volume = "595",
    pages = "109--117",
    year = "2004"
}

@article{E852:2004rfa,
    author = "Lu, M. and others",
    collaboration = "E852",
    title = "{Exotic meson decay to $\omega \pi^0 \pi^-$}",
    eprint = "hep-ex/0405044",
    archivePrefix = "arXiv",
    reportNumber = "JLAB-PHY-04-20",
    doi = "10.1103/PhysRevLett.94.032002",
    journal = "Phys. Rev. Lett.",
    volume = "94",
    pages = "032002",
    year = "2005"
}

@article{Baker:2003jh,
    author = "Baker, C. A. and others",
    title = "{Confirmation of $a_0(1450)$ and $\pi_1(1600)$ in $\bar{p}p \to \omega \pi^+ \pi^- \pi^0$ at rest}",
    doi = "10.1016/S0370-2693(03)00643-9",
    journal = "Phys. Lett. B",
    volume = "563",
    pages = "140--149",
    year = "2003"
}

@article{CLEO:2011upl,
    author = "Adams, G. S. and others",
    collaboration = "CLEO",
    title = "{Amplitude analyses of the decays $\chi_{c1}\to \eta \pi^+ \pi^-$ and $\chi_{c1} \to \eta' \pi^+ \pi^-$}",
    eprint = "1109.5843",
    archivePrefix = "arXiv",
    primaryClass = "hep-ex",
    reportNumber = "CLNS-11-2080, CLEO-11-06",
    doi = "10.1103/PhysRevD.84.112009",
    journal = "Phys. Rev. D",
    volume = "84",
    pages = "112009",
    year = "2011"
}

@article{JPAC:2018zyd,
    author = "Rodas, A. and others",
    collaboration = "JPAC",
    title = "{Determination of the pole position of the lightest hybrid meson candidate}",
    eprint = "1810.04171",
    archivePrefix = "arXiv",
    primaryClass = "hep-ph",
    reportNumber = "JLAB-THY-18-2839",
    doi = "10.1103/PhysRevLett.122.042002",
    journal = "Phys. Rev. Lett.",
    volume = "122",
    number = "4",
    pages = "042002",
    year = "2019"
}

@article{BESIII:2022iwi,
    author = "Ablikim, M. and others",
    collaboration = "BESIII",
    title = "{Partial wave analysis of $J/\psi \to \gamma \eta \eta'$}",
    eprint = "2202.00623",
    archivePrefix = "arXiv",
    primaryClass = "hep-ex",
    doi = "10.1103/PhysRevD.106.072012",
    journal = "Phys. Rev. D",
    volume = "106",
    number = "7",
    pages = "072012",
    year = "2022",
    note = "[Erratum: Phys.Rev.D 107, 079901 (2023)]"
}

@article{BESIII:2022riz,
    author = "Ablikim, M. and others",
    collaboration = "BESIII",
    title = "{Observation of an Isoscalar Resonance with Exotic $J^{PC}=1^{-+}$ Quantum Numbers in $J/\psi \to \gamma \eta \eta'$}",
    eprint = "2202.00621",
    archivePrefix = "arXiv",
    primaryClass = "hep-ex",
    doi = "10.1103/PhysRevLett.129.192002",
    journal = "Phys. Rev. Lett.",
    volume = "129",
    number = "19",
    pages = "192002",
    year = "2022",
    note = "[Erratum: Phys.Rev.Lett. 130, 159901 (2023)]"
}

@article{Zhang:2025xee,
    author = "Zhang, Fu-Yuan and Huang, Qi and Wang, Li-Ming",
    title = "{Spectral analysis and decay mechanisms of $1^{-+}$ hybrid states in light meson sector}",
    eprint = "2503.01443",
    archivePrefix = "arXiv",
    primaryClass = "hep-ph",
    doi = "10.1103/1wkx-3s2l",
    journal = "Phys. Rev. D",
    volume = "113",
    number = "1",
    pages = "014002",
    year = "2026"
}

@article{Ma:2025cew,
    author = "Ma, Zi-Xuan and Huang, Qi and Chen, Rui and Wang, Li-Ming and Tan, Yue and Hu, Xiao-Huang and He, Jun and Huang, Hong-Xia",
    title = "{Proper constituent gluon mass as the final piece to construct hybrid mesons}",
    eprint = "2504.05818",
    archivePrefix = "arXiv",
    primaryClass = "hep-ph",
    doi = "10.1103/yq7l-jmyg",
    journal = "Phys. Rev. D",
    volume = "112",
    number = "11",
    pages = "L111503",
    year = "2025"
}

@article{Zhu:2005hp,
    author = "Zhu, Shi-Lin",
    title = "{The Possible interpretations of $Y(4260)$}",
    eprint = "hep-ph/0507025",
    archivePrefix = "arXiv",
    doi = "10.1016/j.physletb.2005.08.068",
    journal = "Phys. Lett. B",
    volume = "625",
    pages = "212",
    year = "2005"
}

@article{Close:2005iz,
    author = "Close, Frank E. and Page, Philip R.",
    title = "{Gluonic charmonium resonances at BaBar and BELLE?}",
    eprint = "hep-ph/0507199",
    archivePrefix = "arXiv",
    doi = "10.1016/j.physletb.2005.09.016",
    journal = "Phys. Lett. B",
    volume = "628",
    pages = "215--222",
    year = "2005"
}

@article{Zhu:1998sv,
    author = "Zhu, Shi-Lin",
    title = "{Masses and decay widths of heavy hybrid mesons}",
    eprint = "hep-ph/9812405",
    archivePrefix = "arXiv",
    doi = "10.1103/PhysRevD.60.014008",
    journal = "Phys. Rev. D",
    volume = "60",
    pages = "014008",
    year = "1999"
}

@article{Kalashnikova:2016bta,
    author = "Kalashnikova, Yu. S. and Nefediev, A. V.",
    title = "{QCD string in excited heavy-light mesons and heavy-quark hybrids}",
    eprint = "1611.10066",
    archivePrefix = "arXiv",
    primaryClass = "hep-ph",
    doi = "10.1103/PhysRevD.94.114007",
    journal = "Phys. Rev. D",
    volume = "94",
    number = "11",
    pages = "114007",
    year = "2016"
}

@article{Kalashnikova:2008qr,
    author = "Kalashnikova, Yu. S. and Nefediev, A. V.",
    title = "{Spectra and decays of hybrid charmonia}",
    eprint = "0801.2036",
    archivePrefix = "arXiv",
    primaryClass = "hep-ph",
    doi = "10.1103/PhysRevD.77.054025",
    journal = "Phys. Rev. D",
    volume = "77",
    pages = "054025",
    year = "2008"
}

@article{Oncala:2017hop,
    author = "Oncala, Rub{\'e}n and Soto, Joan",
    title = "{Heavy Quarkonium Hybrids: Spectrum, Decay and Mixing}",
    eprint = "1702.03900",
    archivePrefix = "arXiv",
    primaryClass = "hep-ph",
    reportNumber = "ICCUB-17-004, NIKHF-2017-005",
    doi = "10.1103/PhysRevD.96.014004",
    journal = "Phys. Rev. D",
    volume = "96",
    number = "1",
    pages = "014004",
    year = "2017"
}

@article{Govaerts:1984hc,
    author = "Govaerts, J. and Reinders, L. J. and Rubinstein, H. R. and Weyers, J.",
    title = "{Hybrid quarkonia from QCD sum rules}",
    reportNumber = "BONN-HE-84-28",
    doi = "10.1016/0550-3213(85)90609-1",
    journal = "Nucl. Phys. B",
    volume = "258",
    pages = "215--229",
    year = "1985"
}

@article{Govaerts:1985fx,
    author = "Govaerts, J. and Reinders, L. J. and Weyers, J.",
    title = "{Radial Excitations and Exotic Mesons via {QCD} Sum Rules}",
    reportNumber = "UTTG-06-85",
    doi = "10.1016/0550-3213(85)90505-X",
    journal = "Nucl. Phys. B",
    volume = "262",
    pages = "575--592",
    year = "1985"
}

@article{Govaerts:1986pp,
    author = "Govaerts, J. and Reinders, L. J. and Francken, P. and Gonze, X. and Weyers, J.",
    title = "{Coupled {QCD} Sum Rules for Hybrid Mesons}",
    reportNumber = "BUTP-86/14-BERN",
    doi = "10.1016/0550-3213(87)90056-3",
    journal = "Nucl. Phys. B",
    volume = "284",
    pages = "674",
    year = "1987"
}

@article{Qiao:2010zh,
    author = "Qiao, Cong-Feng and Tang, Liang and Hao, Gang and Li, Xue-Qian",
    title = "{Determining $1^{--}$ Heavy Hybrid Masses via QCD Sum Rules}",
    eprint = "1012.2614",
    archivePrefix = "arXiv",
    primaryClass = "hep-ph",
    doi = "10.1088/0954-3899/39/1/015005",
    journal = "J. Phys. G",
    volume = "39",
    pages = "015005",
    year = "2012"
}

@article{Harnett:2012gs,
    author = "Harnett, D. and Kleiv, R. T. and Steele, T. G. and Jin, Hong-ying",
    title = "{Axial Vector $J^{PC}=1^{++}$ Charmonium and Bottomonium Hybrid Mass Predictions with QCD Sum-Rules}",
    eprint = "1206.6776",
    archivePrefix = "arXiv",
    primaryClass = "hep-ph",
    doi = "10.1088/0954-3899/39/12/125003",
    journal = "J. Phys. G",
    volume = "39",
    pages = "125003",
    year = "2012"
}

@article{Chen:2013zia,
    author = "Chen, Wei and Kleiv, R. T. and Steele, T. G. and Bulthuis, B. and Harnett, D. and Ho, J. and Richards, T. and Zhu, Shi-Lin",
    title = "{Mass Spectrum of Heavy Quarkonium Hybrids}",
    eprint = "1304.4522",
    archivePrefix = "arXiv",
    primaryClass = "hep-ph",
    doi = "10.1007/JHEP09(2013)019",
    journal = "JHEP",
    volume = "09",
    pages = "019",
    year = "2013"
}

@article{Wang:2025ypo,
    author = "Wang, Zhi-Gang",
    title = "{Mass spectrum of the hidden-charm hybrid states via QCD sum rules}",
    eprint = "2412.11038",
    archivePrefix = "arXiv",
    primaryClass = "hep-ph",
    doi = "10.1103/hv4x-dnmt",
    journal = "Phys. Rev. D",
    volume = "111",
    number = "11",
    pages = "114009",
    year = "2025"
}

@article{Cheung:2016bym,
    author = "Cheung, Gavin K. C. and O'Hara, Cian and Moir, Graham and Peardon, Michael and Ryan, Sin{\'e}ad M. and Thomas, Christopher E. and Tims, David",
    collaboration = "Hadron Spectrum",
    title = "{Excited and exotic charmonium, $D_s$ and $D$ meson spectra for two light quark masses from lattice QCD}",
    eprint = "1610.01073",
    archivePrefix = "arXiv",
    primaryClass = "hep-lat",
    reportNumber = "DAMTP-2016-63",
    doi = "10.1007/JHEP12(2016)089",
    journal = "JHEP",
    volume = "12",
    pages = "089",
    year = "2016"
}

@article{Juge:2002br,
    author = "Juge, K. Jimmy and Kuti, Julius and Morningstar, Colin",
    title = "{Fine structure of the QCD string spectrum}",
    eprint = "hep-lat/0207004",
    archivePrefix = "arXiv",
    doi = "10.1103/PhysRevLett.90.161601",
    journal = "Phys. Rev. Lett.",
    volume = "90",
    pages = "161601",
    year = "2003"
}

@article{HadronSpectrum:2012gic,
    author = "Liu, Liuming and Moir, Graham and Peardon, Michael and Ryan, Sinead M. and Thomas, Christopher E. and Vilaseca, Pol and Dudek, Jozef J. and Edwards, Robert G. and Joo, Balint and Richards, David G.",
    collaboration = "Hadron Spectrum",
    title = "{Excited and exotic charmonium spectroscopy from lattice QCD}",
    eprint = "1204.5425",
    archivePrefix = "arXiv",
    primaryClass = "hep-ph",
    reportNumber = "TCDMATH-12-04, JLAB-THY-12-1510",
    doi = "10.1007/JHEP07(2012)126",
    journal = "JHEP",
    volume = "07",
    pages = "126",
    year = "2012"
}

@article{Farina:2020slb,
    author = "Farina, Christian and Garcia Tecocoatzi, Hugo and Giachino, Alessandro and Santopinto, Elena and Swanson, Eric S.",
    title = "{Heavy hybrid decays in a constituent gluon model}",
    eprint = "2005.10850",
    archivePrefix = "arXiv",
    primaryClass = "hep-ph",
    doi = "10.1103/PhysRevD.102.014023",
    journal = "Phys. Rev. D",
    volume = "102",
    number = "1",
    pages = "014023",
    year = "2020"
}

@article{Isgur:1984bm,
    author = "Isgur, Nathan and Paton, Jack E.",
    title = "{A Flux Tube Model for Hadrons in QCD}",
    reportNumber = "Print-84-0830 (TORONTO)",
    doi = "10.1103/PhysRevD.31.2910",
    journal = "Phys. Rev. D",
    volume = "31",
    pages = "2910",
    year = "1985"
}

@article{Barnes:1995hc,
    author = "Barnes, Ted and Close, F. E. and Swanson, E. S.",
    title = "{Hybrid and conventional mesons in the flux tube model: Numerical studies and their phenomenological implications}",
    eprint = "hep-ph/9501405",
    archivePrefix = "arXiv",
    reportNumber = "ORNL-CTP-95-02, RAL-94-106",
    doi = "10.1103/PhysRevD.52.5242",
    journal = "Phys. Rev. D",
    volume = "52",
    pages = "5242--5256",
    year = "1995"
}

@article{Close:1994hc,
    author = "Close, Frank E. and Page, Philip R.",
    title = "{The Production and decay of hybrid mesons by flux tube breaking}",
    eprint = "hep-ph/9411301",
    archivePrefix = "arXiv",
    reportNumber = "RAL-94-116, OUTP-94-29-P",
    doi = "10.1016/0550-3213(95)00085-7",
    journal = "Nucl. Phys. B",
    volume = "443",
    pages = "233--254",
    year = "1995"
}

@article{Merlin:1986tz,
    author = "Merlin, John and Paton, Jack E.",
    title = "{Spin Interactions in the Flux Tube Model and Hybrid Meson Masses}",
    reportNumber = "OXFORD-TP 62/86",
    doi = "10.1103/PhysRevD.35.1668",
    journal = "Phys. Rev. D",
    volume = "35",
    pages = "1668",
    year = "1987"
}

@article{Page:1998gz,
    author = "Page, Philip R. and Swanson, Eric S. and Szczepaniak, Adam P.",
    title = "{Hybrid meson decay phenomenology}",
    eprint = "hep-ph/9808346",
    archivePrefix = "arXiv",
    doi = "10.1103/PhysRevD.59.034016",
    journal = "Phys. Rev. D",
    volume = "59",
    pages = "034016",
    year = "1999"
}

@article{Kalashnikova:2002tg,
    author = "Kalashnikova, Yu. S. and Kuzmenko, D. S.",
    title = "{Hybrid adiabatic potentials in the QCD string model}",
    eprint = "hep-ph/0203128",
    archivePrefix = "arXiv",
    reportNumber = "ITEP-2-02",
    doi = "10.1134/1.1577918",
    journal = "Phys. Atom. Nucl.",
    volume = "66",
    pages = "955--967",
    year = "2003"
}

@article{Horn:1977rq,
    author = "Horn, D. and Mandula, J.",
    title = "{A Model of Mesons with Constituent Gluons}",
    reportNumber = "CALT-68-575",
    doi = "10.1103/PhysRevD.17.898",
    journal = "Phys. Rev. D",
    volume = "17",
    pages = "898",
    year = "1978"
}

@article{LeYaouanc:1984gh,
    author = "Le Yaouanc, A. and Oliver, L. and Pene, O. and Raynal, J. C. and Ono, S.",
    title = "{$q \bar{q} g$ Hybrid Mesons in $\psi \to \gamma$ + Hadrons}",
    reportNumber = "LPTHE 84/35",
    doi = "10.1007/BF01575740",
    journal = "Z. Phys. C",
    volume = "28",
    pages = "309--315",
    year = "1985"
}

@article{Swanson:1998kx,
    author = "Swanson, E. S. and Szczepaniak, Adam P.",
    title = "{Heavy hybrids with constituent gluons}",
    eprint = "hep-ph/9804219",
    archivePrefix = "arXiv",
    doi = "10.1103/PhysRevD.59.014035",
    journal = "Phys. Rev. D",
    volume = "59",
    pages = "014035",
    year = "1999"
}

@article{Szczepaniak:2006nx,
    author = "Szczepaniak, Adam P. and Krupinski, Pawel",
    title = "{Energy spectrum of the low-lying gluon excitations in the Coulomb gauge}",
    eprint = "hep-ph/0604098",
    archivePrefix = "arXiv",
    doi = "10.1103/PhysRevD.73.116002",
    journal = "Phys. Rev. D",
    volume = "73",
    pages = "116002",
    year = "2006"
}

@article{Guo:2008yz,
    author = "Guo, Peng and Szczepaniak, Adam P. and Galata, Giuseppe and Vassallo, Andrea and Santopinto, Elena",
    title = "{Heavy quarkonium hybrids from Coulomb gauge QCD}",
    eprint = "0807.2721",
    archivePrefix = "arXiv",
    primaryClass = "hep-ph",
    doi = "10.1103/PhysRevD.78.056003",
    journal = "Phys. Rev. D",
    volume = "78",
    pages = "056003",
    year = "2008"
}

@book{doi:10.1142/0270,
author = {Varshalovich, D A and Moskalev, A N and Khersonskii, V K},
title = {Quantum Theory of Angular Momentum},
publisher = {WORLD SCIENTIFIC},
year = {1988},
doi = {10.1142/0270},
address = {},
edition   = {},
URL = {https://www.worldscientific.com/doi/abs/10.1142/0270},
eprint = {https://www.worldscientific.com/doi/pdf/10.1142/0270}
}

@article{Chen:2025pvk,
    author = "Chen, Bing and Liu, Xiang",
    title = "{Investigating hybrid mesons with $0^{+-}$ and $2^{+-}$ exotic quantum numbers}",
    eprint = "2503.06116",
    archivePrefix = "arXiv",
    primaryClass = "hep-ph",
    doi = "10.1140/epjc/s10052-025-14509-y",
    journal = "Eur. Phys. J. C",
    volume = "85",
    number = "7",
    pages = "788",
    year = "2025"
}

@article{Barnes:1982tx,
    author = "Barnes, Ted and Close, F. E. and de Viron, F.",
    title = "{$\mathrm{Q\bar{Q}G}$ Hermaphrodite Mesons in the MIT Bag Model}",
    reportNumber = "RL-82-088",
    doi = "10.1016/0550-3213(83)90004-4",
    journal = "Nucl. Phys. B",
    volume = "224",
    pages = "241",
    year = "1983"
}

@article{Chanowitz:1982qj,
    author = "Chanowitz, Michael S. and Sharpe, Stephen R.",
    title = "{Hybrids: Mixed States of Quarks and Gluons}",
    reportNumber = "LBL-14865",
    doi = "10.1016/0550-3213(83)90635-1",
    journal = "Nucl. Phys. B",
    volume = "222",
    pages = "211--244",
    year = "1983",
    note = "[Erratum: Nucl.Phys.B 228, 588--588 (1983)]"
}

@article{Agaev:2025llz,
    author = "Agaev, S. S. and Azizi, K. and Sundu, H.",
    title = "{Tensor hybrid charmonia}",
    eprint = "2504.03003",
    archivePrefix = "arXiv",
    primaryClass = "hep-ph",
    doi = "10.1103/7g78-zk74",
    journal = "Phys. Rev. D",
    volume = "112",
    number = "1",
    pages = "014003",
    year = "2025"
}

@article{Braaten:2014qka,
    author = "Braaten, Eric and Langmack, Christian and Smith, D. Hudson",
    title = "{Born-Oppenheimer Approximation for the $XYZ$ Mesons}",
    eprint = "1402.0438",
    archivePrefix = "arXiv",
    primaryClass = "hep-ph",
    doi = "10.1103/PhysRevD.90.014044",
    journal = "Phys. Rev. D",
    volume = "90",
    number = "1",
    pages = "014044",
    year = "2014"
}

@article{Berwein:2015vca,
    author = "Berwein, Matthias and Brambilla, Nora and Tarr{\'u}s Castell{\`a}, Jaume and Vairo, Antonio",
    title = "{Quarkonium Hybrids with Nonrelativistic Effective Field Theories}",
    eprint = "1510.04299",
    archivePrefix = "arXiv",
    primaryClass = "hep-ph",
    reportNumber = "TUM-EFT-45-14",
    doi = "10.1103/PhysRevD.92.114019",
    journal = "Phys. Rev. D",
    volume = "92",
    number = "11",
    pages = "114019",
    year = "2015"
}

@article{Brambilla:2018pyn,
    author = "Brambilla, Nora and Lai, Wai Kin and Segovia, Jorge and Tarr{\'u}s Castell{\`a}, Jaume and Vairo, Antonio",
    title = "{Spin structure of heavy-quark hybrids}",
    eprint = "1805.07713",
    archivePrefix = "arXiv",
    primaryClass = "hep-ph",
    reportNumber = "TUM-EFT-95-17, TUM-EFT 95/17",
    doi = "10.1103/PhysRevD.99.014017",
    journal = "Phys. Rev. D",
    volume = "99",
    number = "1",
    pages = "014017",
    year = "2019",
    note = "[Erratum: Phys.Rev.D 101, 099902 (2020)]"
}

@article{Brambilla:2019jfi,
    author = "Brambilla, Nora and Lai, Wai Kin and Segovia, Jorge and Tarr{\'u}s Castell{\`a}, Jaume",
    title = "{QCD spin effects in the heavy hybrid potentials and spectra}",
    eprint = "1908.11699",
    archivePrefix = "arXiv",
    primaryClass = "hep-ph",
    reportNumber = "TUM-EFT 129/19",
    doi = "10.1103/PhysRevD.101.054040",
    journal = "Phys. Rev. D",
    volume = "101",
    number = "5",
    pages = "054040",
    year = "2020"
}

@article{Soto:2023lbh,
    author = "Soto, Joan and Valls, Sandra Tom{\`a}s",
    title = "{Hyperfine splittings of heavy quarkonium hybrids}",
    eprint = "2302.01765",
    archivePrefix = "arXiv",
    primaryClass = "hep-ph",
    doi = "10.1103/PhysRevD.108.014025",
    journal = "Phys. Rev. D",
    volume = "108",
    number = "1",
    pages = "014025",
    year = "2023"
}

@article{Page:1996rj,
    author = "Page, Philip R.",
    title = "{Why hybrid meson coupling to two $S$-wave mesons is suppressed}",
    eprint = "hep-ph/9611375",
    archivePrefix = "arXiv",
    reportNumber = "MC-TH-96-26",
    doi = "10.1016/S0370-2693(97)00438-3",
    journal = "Phys. Lett. B",
    volume = "402",
    pages = "183--188",
    year = "1997"
}

@book{Peskin:1995ev,
    author = "Peskin, Michael E. and Schroeder, Daniel V.",
    title = "{An Introduction to quantum field theory}",
    doi = "10.1201/9780429503559",
    isbn = "978-0-201-50397-5, 978-0-429-50355-9, 978-0-429-49417-8",
    publisher = "Addison-Wesley",
    address = "Reading, USA",
    year = "1995"
}

@article{Close:2003mb,
    author = "Close, F. E. and Godfrey, Stephen",
    title = "{Charmonium hybrid production in exclusive B meson decays}",
    eprint = "hep-ph/0305285",
    archivePrefix = "arXiv",
    reportNumber = "ADP-03-120-T558, TRI-PP-03-04, OUTP-03-13-P",
    doi = "10.1016/j.physletb.2003.09.011",
    journal = "Phys. Lett. B",
    volume = "574",
    pages = "210--216",
    year = "2003"
}

@article{ParticleDataGroup:2024cfk,
    author = "Navas, S. and others",
    collaboration = "Particle Data Group",
    title = "{Review of particle physics}",
    doi = "10.1103/PhysRevD.110.030001",
    journal = "Phys. Rev. D",
    volume = "110",
    number = "3",
    pages = "030001",
    year = "2024"
}

@article{BESIII:2024ths,
    author = "Ablikim, Medina and others",
    collaboration = "BESIII",
    title = "{Precise Measurement of Born Cross Sections for $e^+e^- \to D\bar{D}$ at $\sqrt{s}=3.80-4.95\,\text{GeV}$}",
    eprint = "2402.03829",
    archivePrefix = "arXiv",
    primaryClass = "hep-ex",
    doi = "10.1103/PhysRevLett.133.081901",
    journal = "Phys. Rev. Lett.",
    volume = "133",
    number = "8",
    pages = "081901",
    year = "2024"
}

@article{Wang:2025dur,
    author = "Wang, Xiongfei and Liu, Xiang and Gao, Yuanning",
    title = "{Colloquium: Hadron production in open-charm meson pairs at $e^+e^-$ colliders}",
    eprint = "2502.15117",
    archivePrefix = "arXiv",
    primaryClass = "hep-ex",
    doi = "10.1103/2mrp-chly",
    journal = "Rev. Mod. Phys.",
    volume = "98",
    number = "2",
    pages = "021001",
    year = "2026"
}

\end{document}